\documentclass[journal]{IEEEtran}
\usepackage{graphicx}
\usepackage{amsmath}
\usepackage{amssymb}
\usepackage{booktabs}
\usepackage{amsfonts}
\usepackage{arydshln}
\usepackage{dsfont}
\usepackage{algorithm}
\usepackage{multirow}
\usepackage{cite}
\usepackage{algpseudocode} 
\usepackage{xcolor}
\PassOptionsToPackage{hyphens}{url}
\usepackage{hyperref}

\usepackage[capitalize]{cleveref}
\crefname{section}{Sec.}{Secs.}
\Crefname{section}{Section}{Sections}
\Crefname{table}{Table}{Tables}
\crefname{table}{Tab.}{Tabs.}

\begin{document}

\title{Differential Analysis of Triggers and Benign Features for Black-Box DNN Backdoor Detection}

\author{Hao~Fu,
        Prashanth~Krishnamurthy, Member, IEEE
        Siddharth~Garg, Member, IEEE,
        Farshad~Khorrami, Senior Member, IEEE
\thanks{Department
of Electrical and Computer Engineering, New York University, Brooklyn, NY, 11201, USA. E-mail: \{hf881, prashanth.krishnamurthy, sg175, khorrami\} @nyu.edu. This work was supported in part by the Army Research Office under grant number W911NF-21-1-0155 and by the New York University Abu Dhabi (NYUAD) Center for Artificial Intelligence and Robotics, funded by Tamkeen under the NYUAD Research Institute Award CG010.  Code is available at \url{https://github.com/fu1001hao/Five-Metrics-Detector.git}.}
}

\maketitle

\begin{abstract}
   This paper proposes a data-efficient detection method for deep neural networks against  backdoor attacks under a black-box scenario. The proposed approach is motivated by the intuition that features corresponding to  triggers have a higher influence in determining the backdoored network output than any other benign features. To quantitatively measure the effects of triggers and benign features on determining the backdoored network output, we introduce five metrics.  To calculate the five-metric values for a given input, we first generate several synthetic samples by injecting the input's partial contents into clean validation samples. Then, the five metrics are computed by using the output labels of the corresponding synthetic samples. One contribution of this work is the use of a tiny clean validation dataset.
   Having the computed five metrics, five novelty detectors are trained from the validation dataset. A meta novelty detector fuses the output of the five trained novelty detectors to generate a meta confidence score. During online testing, our method determines if online samples are poisoned or not via assessing their meta confidence scores output by the meta novelty detector.  We show the efficacy of our methodology through a broad range of backdoor attacks, including ablation studies and comparison to existing approaches. Our methodology is promising since the proposed five metrics quantify the inherent differences between clean and poisoned samples. Additionally, our detection method can be incrementally improved by appending more metrics that may be proposed to address future advanced attacks. 
\end{abstract}


\section{Introduction}
\label{sec:intro}
 Deep neural networks (DNN) should be secure and  reliable since they are utilized in many applications \cite{SWT14,TYRW14,CSKX15,FKK20,PSCKK18,PKGK19,PCKK19,DKPK21,PSCKK20,PKGK21,DKPK23}. Therefore, studying the security problems for DNN is an important research topic \cite{AND18,CW17,EEFLRXPKS18,GSS14,HWCCS17,MSFFF17, FVKGK20}. This paper considers defending against  backdoor attacks in  neural networks for classification tasks under a black-box scenario,  {in which only the network output is accessible and other information (e.g., model weights and intermediate layer outputs) is not available.} Backdoor attacks may appear in models trained by a third party. In backdoor attacks, the attacker injects   triggers into the network during the training phase. During the testing phase, the backdoored neural network outputs the attacker-chosen labels whenever the corresponding triggers appear.    This paper proposes a novel defense against backdoor attacks based on a differential analysis of the behaviors of backdoor triggers and benign features used by the network for classification.

Although there is substantial literature on backdoor attacks and defenses,  effective detection methods are scarce. Detecting backdoors  is challenging due to the  asymmetric-information advantage available to the adversary (i.e., the attacker has complete control of the trigger, whereas the defender has  little information about the trigger). Among the existing defenses, many  make assumptions about the trigger, such as assuming that the trigger is small or non-adaptive. However, the assumptions may not be valid in  real-world situations because a clever attacker can design any trigger. Another set of existing literature assumes that the defender has access to a contaminated dataset that contains trigger information. In real-world cases, accessing a contaminated dataset may not be feasible for the defender. Some studies do not have assumptions on triggers or a need for contaminated datasets. However, they require a large amount of clean data. Collecting a large amount of clean data may not  always be affordable to the defender. Many works focus on detecting if a neural network is backdoored and should be abandoned. However, this paper is interested in designing a detection algorithm with limited clean validation samples to reject poisoned inputs (i.e., the inputs with triggers) so that the backdoored model can still be used without causing considerable loss. Therefore, we propose a detection method that has no assumptions on triggers, does not require the availability of a contaminated dataset, and only requires a tiny clean validation dataset.

Our method is inspired by the definition of the backdoor attack: the attacker controls the neural network output by overriding the original logic of the network when he/she presents triggers. This overriding behavior of the neural network will be exposed only for poisoned inputs, whereas for clean inputs, the neural network behaves normally. Therefore, we claim that the trigger has a higher influence than the benign features in determining the backdoored network output. Based on this difference, we propose five metrics (i.e., robustness $r$,  weakness $w$,  sensitivity $s$,  inverse sensitivity $is$, and  noise invariance $Inv$) such that a function exists to separate the clean and poisoned inputs regarding their five-metric values.

To calculate the five-metric values for a given input, our detection algorithm generates a few synthetic images by injecting the input's partial contents into samples from a tiny validation dataset and utilizes the output labels corresponding to the synthetic images.  Having the computed five metrics, five novelty detectors are trained from the tiny validation dataset. The trained novelty detectors will output high (resp., low) confidence scores for  clean (resp., poisoned) inputs because a clean (resp., poisoned) input's metric values will be similar to (resp., different from) the clean validation samples' metric values (i.e., the training data for the novelty detectors) with a high probability.  Thereafter, a meta novelty detector fuses the confidence scores by the five trained novelty detectors. During online testing, our method determines if a new input is poisoned and should be rejected via assessing its meta confidence score output by the meta novelty detector. 

{Besides the backdoor study, the proposed five metrics also contribute to solving two important problems for hand-crafted features-based anomaly detection \cite{CYL11,MOS09,KN09}: 1) designing an effective descriptor, and 2) deciding suitable features for specific anomaly detection situations \cite{FYL17}.  Our approach is novel in that it contributes effective hand-crafted features for backdoor detection (which can be regarded as an anomaly detection) using the proposed five novel metrics and achieves high accuracy under a black-box scenario with scarce available clean samples, whereas existing hand-crafted features-based anomaly detection methods are not designed for backdoor detection purposes and hence are not as effective as ours.} 

Overall, the contributions of this paper include: 
\begin{itemize}
    \item We utilize the conceptual ways that triggers can be injected into a backdoored  network in order to achieve an unsupervised approach for backdoor detection;
    \item We propose five metrics to measure the behavior of the network for different scenarios; 
    \item We propose a data-efficient  online detection algorithm using the five-metric values as inputs to detect poisoned inputs for   neural networks under a black-box scenario;
    \item We evaluate and compare the efficacy of our detection approach with other  methods on various backdoor attacks.
\end{itemize}

\section{Related Work}
\label{sec:rela}

{\bf Backdoor attack} was first proposed by \cite{GLDG19} and  \cite{LMALZWZ17}.  Several types of backdooring triggers have been studied including triggers with semantic real-world meaning \cite{WPBYZZ21}, hidden invisible triggers \cite{LLWLHL21, SSP20, LXZZZ20, TA21}, smooth triggers \cite{ZPMJ21}, and reflection triggers \cite{LMBL20}. Backdoor attacks have been devised in several contexts including federated learning \cite{XHCL19,BVHES20,AMMK21}, transfer learning \cite{YLZZ19}, graph networks \cite{ZJWG21},  text-classification \cite{DCL19},  and out-sourced cloud environments \cite{GCWHMSZ21}. Several scenarios/variants of backdoor attacks have been considered including  all-label attacks \cite{GLDG19}, clean label attacks \cite{ZMZBCJ20,ltkg20}, and defense-aware attacks \cite{LDG18}. The backdooring mechanism has also been applied for benign/beneficial purposes, such as watermarking for patent protection \cite{SLWLK21}.

{\bf Backdoor attack defenses} can be classified into several groups regarding their assumptions or proposed methods. The reverse-engineering-based approaches \cite{WYSLVZZ19,QYL19,LLTMAZ19, GWXDS19} attempt to solve an optimization problem under certain restrictive assumptions; hence, those methods are effective only in a small portion of cases. \cite{DYDPXSZ21} proposed a gradient-free technology to reverse-engineer the trigger with limited data. Clustering-based approaches \cite{TLM18,CCBLELMS18,XMK21,HK20} assume a contaminated training dataset is available, whose acquisition might be expensive. Novelty-detection-based approaches \cite{LLLS18,APCC19,ONP20,LLMZSL18,CFZK19,ZNWXW20} require enough  clean validation samples for training the complex novelty detector models, especially for the neural-network-based novelty detectors. The retraining-based approaches \cite{K20,FVKGK20} also need a reasonable amount of clean data to achieve high performance. If the available clean data is not sufficient, their performance degrades dramatically. \cite{FVKGK22} used online data to improve detection accuracy. However, the method becomes ineffective when online data is limited. Some works tested if a network has a backdoor \cite{HPJT20, LLTMAZ19} and  should be abandoned.  Fine-pruning \cite{LDG18} and STRIP \cite{GXWCRN19} have their assumptions and limitations that require further improvements. Some works modify the original problem and show the behavior of the backdoored networks in their settings, such as noise response analysis \cite{ETWM21} and generation of universal litmus patterns \cite{KSPH20}. \cite{ZPMJ21} studied backdoor attacks in the frequency domain.

\section{Problem Formulation}
\label{sec:prob}

\subsection{Background and Assumptions}

\begin{table}
\caption{Behavior of benign and backdoored neural networks.}
    \centering
    \begin{tabular}{c|cc}
    \hline
     data / model  & Benign   $f$ & Backdoored $f^*$  \\
        \hline
      Clean   $(z, l)$ & $\mathbb{P}(f(z) = l)$ high & $\mathbb{P}(f^*(z) = l)$ high \\
      Poisoned   $(z^*, l^*)$ & $\mathbb{P}(f(z^*) = l)$ high & $\mathbb{P}(f^*(z^*) = l^*)$ high \\
        \hline
    \end{tabular}
    \label{tab:behavior}
\end{table}

The difference between a backdoored network and a benign network in classification is shown in Table~\ref{tab:behavior}: a backdoored network $f^*$ outputs ground truth label $l$ for a clean input $z$ and a wrong label $l^*$ (called attacker-chosen label) for a poisoned input $z^*$ with  a high probability, as shown in the last column in the table. However, poisoning an input adds negligible influence on the output of a benign model $f$, as shown in the middle column in the table.

We assume that only the network output is available to our approach. The network's other information (e.g., gradients, weights, and hidden layer outputs) is not available. This {\bf black-box} assumption makes our detection approach realistic since such internal access into the network may be unavailable due to proprietary/security considerations in real-world cases.  Additionally, this black-box setting is widely used in the literature of neural network backdoor studies   \cite{DYDPXSZ21}. We assume that there is a {\bf small set of clean data} $\{x_i\}_{i=1}^n$ (e.g., with size $n\le 30$) to confirm the performance of $f$ on clean data. We assume that only the backdoor attack appears in our problem. Other types of attacks are out of the scope of this paper.

\subsection{Problem Formulation}
Given a black-box network $f$ and a tiny validation dataset $\{x_i\}_{i=1}^n$, we want to find a detection algorithm $g(\cdot;\{x_i\}_{i=1}^n,f)$ such that $g(z^*;\{x_i\}_{i=1}^n,f) = 1$ with a high probability for poisoned inputs $z^*$ (if $f$ is backdoored) and $g(z;\{x_i\}_{i=1}^n,f)=0$ with a high probability for clean inputs $z$, i.e.,
\begin{align}
    \mathbb{P}(g(z;\{x_i\}_{i=1}^n,f)= 0 | z \text{ is clean})&\ge 1-\epsilon_1, \\
    \mathbb{P}(g(z^*;\{x_i\}_{i=1}^n,f)= 1 | z^* \text{ is poisoned})&\ge 1-\epsilon_2
\end{align} where $\epsilon_{1}$ and $\epsilon_{2}$ are two small positive numbers.

\subsection{Important Concepts}
{\bf Classification Accuracy (CA)} is the ratio of the number of clean inputs for which the network outputs ground-truth labels to the total number of clean inputs. Both backdoored networks and benign networks should have high CA. {\bf Attack Success Rate (ASR)} is the ratio of the number of poisoned inputs for which the network outputs attacker-chosen labels to the total number of poisoned inputs. ASR should be high for backdoored networks and low for benign networks.

{\bf True-Positive Rate (TPR)} is the ratio of the number of  poisoned inputs detected by the detection algorithm to the total number of poisoned inputs. An accurate detection algorithm should have high TPR. {\bf False-Positive Rate (FPR)} is the ratio of the number of clean inputs  misidentified as poisoned by the detection algorithm to the total number of clean inputs. An accurate detection algorithm should  have  low FPR.

{\bf Receiver Operating Characteristic Curve (ROC)} is a graph that shows the detection algorithm's performance at all  thresholds. Its two parameters are TPR and FPR. {\bf Area Under the ROC Curve (AUROC)} is the entire two-dimensional area underneath the entire ROC curve. An accurate detection algorithm should have AUROC close to 1.  {{\bf Area Under the Precision and Recall (AUPR)} is similar to AUROC but with precision\footnote{Precision = TruePositives/(TruePositives+FalsePositives).} and recall\footnote{Recall = TruePositives/(TruePositives+FalseNegatives).}  as its two parameters. AUPR is useful when the testing dataset is imbalanced. Higher AUPR implies a better performance of the approach.}

{\bf Novelty Detector} is a one-class detector that learns the training data distribution and detects if an incoming new sample belongs to this distribution or not. In this paper, clean validation data and clean online data belong to the same distribution, whereas clean validation data and poisoned online data belong to different  distributions. 

\section{Methodology}
\label{sec:meth}

\subsection{Intuition -- Rethinking the Pattern-Based Triggers}
\begin{figure}
    \centering
    \includegraphics[scale=0.25]{ 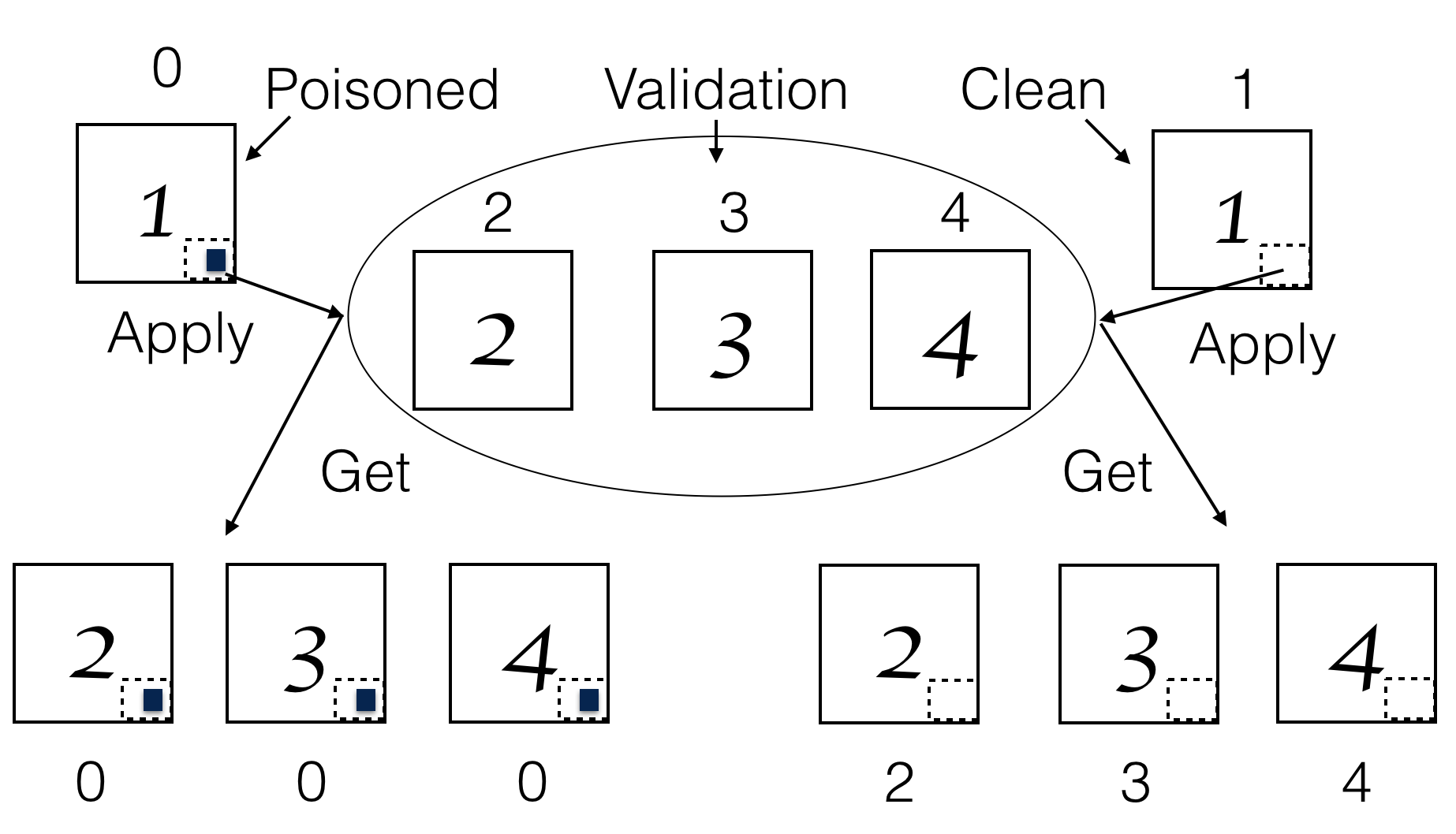}
    \caption{Network outputs before and after pasting the dashed area into the clean validation samples for the poisoned and clean input cases. The trigger is the black pixel. The attacker-chosen label is $0$. The number inside each image is the ground-truth label, whereas the number outside each image is the corresponding backdoored network output label.}
    \label{fig:naive}
\end{figure}

Consider the naive trigger with functionality shown in Fig.~\ref{fig:naive}: the trigger is one pixel with a fixed value located at the lower-right corner of the image. Any image attached with this trigger  makes the backdoored network $f^*$ output the attacker-chosen label $l^*$ (i.e., 0). This shows that the trigger pattern has a higher influence than other benign features in deciding the network output. We measure this influence with the following steps: 1) given an image, we copy its partial content (i.e., the dashed area in Fig.~\ref{fig:naive}) and paste the content into different clean validation samples in the exact corresponding location to generate synthetic images. 2) We feed these synthetic images into the network and observe the outputs. If the image is poisoned and the pasted content contains the trigger, then all the output labels should be $l^*$ (i.e., the left ``Apply-Get'' in Fig.~\ref{fig:naive}). If the image is clean, it is less likely that all the output labels are the same (i.e., the right ``Apply-Get'' in Fig.~\ref{fig:naive}). Therefore, the consistency among the network output labels for synthetic images can be used to measure this influence.  

Based on motivations analogous to the above discussion, we propose five metrics to quantitatively measure the effect of regions of a given image. Using these five metrics as a five-metric set, we will train a classifier that will enable testing of the given image for the presence of triggers.

 \subsection{The Five Metrics}
 
  The five proposed metrics are  robustness $r$,  weakness $w$,  sensitivity $s$,  inverse sensitivity $is$, and  noise invariance $Inv$. Defining them will use the following notations:
  \begin{itemize}
  \item $z$ represents an input image to be evaluated for backdoor presence.
      \item $\{x_i\}_{i=1}^n$ represents $n$ clean validation samples\footnote{ {We require $z$ and $\{x_i\}_{i=1}^n$ to belong to the same domain. For instance, they can be both MNIST-like images for the MNIST dataset}}.
       {
      \item $U_{(\cdot)}$ represents the partial content of the image $(\cdot)$. For example, $U_{z}$ represents the partial content of image $z$.
      \item $paste(\cdot, *)$  pastes $(\cdot)$ into $(*)$ in the exact corresponding location and returns the synthetic image. For example, $ paste(U_z, x_i)$  pastes  $U_z$ into $x_i$ in the exact corresponding location and returns the synthetic image.
       }
      \item  $\mathds{1}$ represents the indicator function\footnote{With $A$ being a set, $\mathds{1}_A(x)=1$ if $x\in A$, and $\mathds{1}_A(x)=0$, otherwise.}.
      \item $f^*$ represents the neural network (possibly backdoored).
      \item $\epsilon \sim \mathcal{N}(0,\delta)$ represents the normal noise tensor.
  \end{itemize}

 {\bf Robustness $r$} quantifies the likelihood that $U_z$ overrides the prediction of the backdoored network $f^*$ on $x_i$:
    \begin{align}
        r = \frac{1}{n}\sum_{i=1}^n \mathds{1} \{f^*(z) = f^*(paste(U_z, x_i))\}.
        \label{robustness}
    \end{align}
    {As one example scenario, if $z$ is poisoned and $U_z$  does not include the benign features of $z$ but includes the trigger, $r$ will be high. Inversely, if $z$ is clean and $U_z$ does not contain the benign features, $r$ will be low. }

{\bf Weakness $w$} quantifies the likelihood that $U_z$ fails to make the backdoored network $f^*$ change its prediction on $x_i$:
    \begin{align}
        w = \frac{1}{n} \sum_{i=1}^n \mathds{1}\{ f^*(x_i) = f^*(paste(U_z, x_i))\}.
        \label{weakness}
    \end{align}   {As one example scenario, if $z$ is poisoned and $U_z$ does not include the benign features of $z$ but includes the trigger, $w$ will be low. Inversely, if $z$ is clean and $U_z$ does not include the benign features,  $w$ will be high. }

{\bf Sensitivity $s$} quantifies the likelihood that $z$ still contains high influence features after $U_{x_i}$ is pasted:
    \begin{align}
        s = \frac{1}{n} \sum_{i=1}^n \mathds{1}\{f^*(z) = f^*(paste(U_{x_i}, z))\}.
        \label{sensitivity}
    \end{align}   {As one example case, if $z$ is poisoned and $U_{x_i}$ contains the benign features of $x_i$ but $paste(U_{x_i}, z)$ still contains the trigger, $s$ will be high. Inversely, if $z$ is clean and $U_{x_i}$ contains the benign features of $x_i$, $s$ will be low. }

{\bf Inverse Sensitivity $is$} quantifies the likelihood that $z$ does not contain high influence features after $U_{x_i}$ is pasted:
    \begin{align}
        is = \frac{1}{n} \sum_{i=1}^n \mathds{1}\{ f^*(x_i) = f^*(paste(U_{x_i}, z))  \}.
        \label{inverse}
    \end{align}   {As one example scenario, if $z$ is poisoned and $U_{x_i}$ contains the benign features of $x_i$ but $paste(U_{x_i}, z)$ still contains the trigger, $is$ will be low. Inversely, if $z$ is clean and  $U_{x_i}$  contains the benign features of $x_i$, $is$ will be high. Thus, the  metrics help distinguish between clean and poisoned samples.}

  {
The following observations are made:
\begin{itemize}
    \item Each metric is expected to contribute in different  ways to detect various triggers, although it is possible that multiple metrics  capture the same trigger in some cases. Fusing all metrics further enhances the true positive rates and detection of the triggers. For example, one can easily design counterexamples for poisoned samples to evade detection using $r$ or $s$. However, $w$ and $is$ help complement $r$ and $s$ to detect those counterexamples. 
    \item Having some reasonable regions $U$ is the key step to distinguishing the clean and poisoned samples. Therefore, we use several  regions. 
    \item These four metrics consider insertions of regions of a given image into corresponding regions of the validation set images (or vice versa), which are expected to be most relevant for pattern-based triggers (i.e., triggers contained in some regions in the image). However, non-pattern-based triggers exist (i.e., triggers that are based on inserting subtle variations throughout the image) and are usually entangled with benign features, which cannot be separated from benign features by any region. Consequently, the performance of these four metrics can be very low on certain non-pattern-based triggers. Therefore, the fifth metric $Inv$ is needed.
\end{itemize}
 }

\begin{figure}
    \centering
    \includegraphics[width=\linewidth]{ 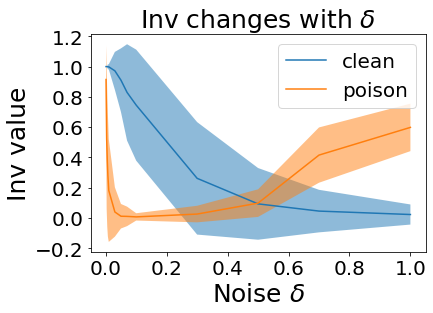}
    \caption{As noise variance $\delta$ changes, $Inv$ for the clean inputs decreases, whereas $Inv$ for the poisoned inputs decreases first and then increases. The plots show the average of multiple runs utilizing multiple samples in each run.  { The involved dataset is GTSRB \cite{SSSI12} and the used trigger is Wanet \cite{TA21}.}}
    \label{fig:inv}
\end{figure}

{\bf Noise Invariance $Inv$} quantifies the robustness of the features in $z$ against noise perturbation:
    \begin{align}
        Inv = \frac{1}{n} \sum_{i=1}^n \mathds{1}\{ f^*(z) = f^*(z+\epsilon_i)\}.
        \label{noise}
    \end{align}   {If $z$ is poisoned and $\epsilon_i$ does not break the function of the benign features of $z$ but breaks the function of the trigger, $Inv$ will be low. Inversely, if $z$ is clean and $\epsilon_i$ does not break the function of the benign features, $Inv$ will be high. Thus, clean and poisoned samples are distinguishable.}
  {
Similarly, finding the ideal noise $\epsilon$ is the key to distinguishing clean and poisoned samples. Therefore, we utilize a pool of noise distributions (i.e., different $\delta$).  Fig.~\ref{fig:inv} is made with noise $\epsilon$ sampled from $\mathcal{N}(0, \delta)$ with different $\delta$ shown in the X-axis. The noise pattern is global with the same shape as the input image (i.e., a noise perturbation $\epsilon\sim\mathcal{N}(0, \delta)$ is added into each pixel of the image). From the figure, the clean and poisoned samples are empirically distinguishable. We also empirically observed that although $Inv$ is designed for non-pattern-based triggers, it also works well on some pattern-based triggers. The explanation is that some generated noise perturbations break the trigger pattern's function but do not break the benign features' influence. Overall, $Inv$ helps in distinguishing clean and poisoned samples, especially for non-pattern-based triggers and sometimes for pattern-based triggers.
}

\subsection{The Pool of Feature Extraction Regions and Noise Variance}
  {
Separating different triggers and benign features could require multiple regions. Therefore, one should use a pool of regions to separate the triggers and benign features.} Similarly, as shown in Fig.~\ref{fig:inv}, a pool of $\delta$ can better capture the poisoned samples. This paper uses a pool of 16 central regions and a pool of 16 noise variances. Specifically, the aspect ratio\footnote{The aspect ratio is the ratio of the height of the central region to the height of the original image.} of the central regions to the image is 0.1, 0.15, 0.2, 0.25, 0.3, 0.35, 0.4,0.45, 0.5,0.55, 0.6,0.65, 0.7,0.75, 0.8, and 0.9, respectively. And the noise variances are 0.001, 0.003, 0.005, 0.007,  0.01,  0.03, 0.05, 0.07, 0.1, 0.3, 0.5, 0.7, 1, 1.5, 2, and 3, respectively. Alg.~\ref{alg:calculating} shows the extraction process used in this paper. We use the aspect ratio to calculate the coordinates of the extraction regions and then copy the content of the regions (i.e., $U_z$ and $U_{x_i}$) to calculate the five-metric values. Since the pool has 16 elements, each metric value will be a 16-dimensional vector. Note that the pools can be expanded by adding more regions and noise.
 {For example, one can use and add extra regions with different locations and shapes. The paper later shows the performance of our approach using central regions and additional corner regions. Similarly, one can use and add extra noise perturbations from a wide range of distributions.} 

 \begin{algorithm}
    \caption{MetricCal($z$, $\{x_i\}_{i=1}^n$)}  
 \label{alg:calculating} 
 \begin{algorithmic}
 \State {\bf Initialization:}
     \State Ratio $\leftarrow$ [0.1, 0.15, 0.2, 0.25, 0.3, 0.35, 0.4,0.45, 0.5,0.55, 0.6,0.65, 0.7,0.75, 0.8, 0.9]
     \State Variance $\leftarrow$ [0.001, 0.003, 0.005, 0.007,  0.01,  0.03, 0.05, 0.07, 0.1, 0.3, 0.5, 0.7, 1, 1.5, 2, 3]

  \State \#\#\# Calculating the Five-Metric Values for $z$ \#\#\#
     \State  {H}, W $\leftarrow$   {Height} of $z$, Width of $z$
     \State   {$r, w, s, is, Inv \leftarrow \{\}, \{\}, \{\}, \{\}, \{\}$}
     \For {$j = 1 \to 16$}
        \State  k $\leftarrow$ Ratio[j]
         \State $m_1, n_1, m_2, n_2 \leftarrow \frac{1-k}{2}H, \frac{1-k}{2}W, \frac{1+k}{2}H, \frac{1+k}{2}W$
         \State $U_{z} \leftarrow z[m_1:m_2, n_1:n_2]$
         \State $\{U_{x_i}\}_{i=1}^n \leftarrow \{x_i[m_1:m_2, n_1:n_2 ]\}_{i=1}^n$
         \State $\delta \leftarrow$ Variance[j]
         \State $r_j^\prime, w_j^\prime, s_j^\prime, is_j^\prime, Inv_j^\prime \leftarrow$ \eqref{robustness}, \eqref{weakness}, \eqref{sensitivity}, \eqref{inverse}, \eqref{noise}
           {
         \State $r.append(r_j^\prime)$,  $w.append(w_j^\prime)$
         \State $s.append(s_j^\prime)$, $is.append(is_j^\prime)$
         \State $Inv.append(Inv_j^\prime)$}
     \EndFor
     \State {\bf Return:} $r, w, s, is, Inv$
 \end{algorithmic}
 \end{algorithm}

\subsection{Novelty Detection Process}
\begin{figure}
    \centering
    \includegraphics[scale=0.27]{ 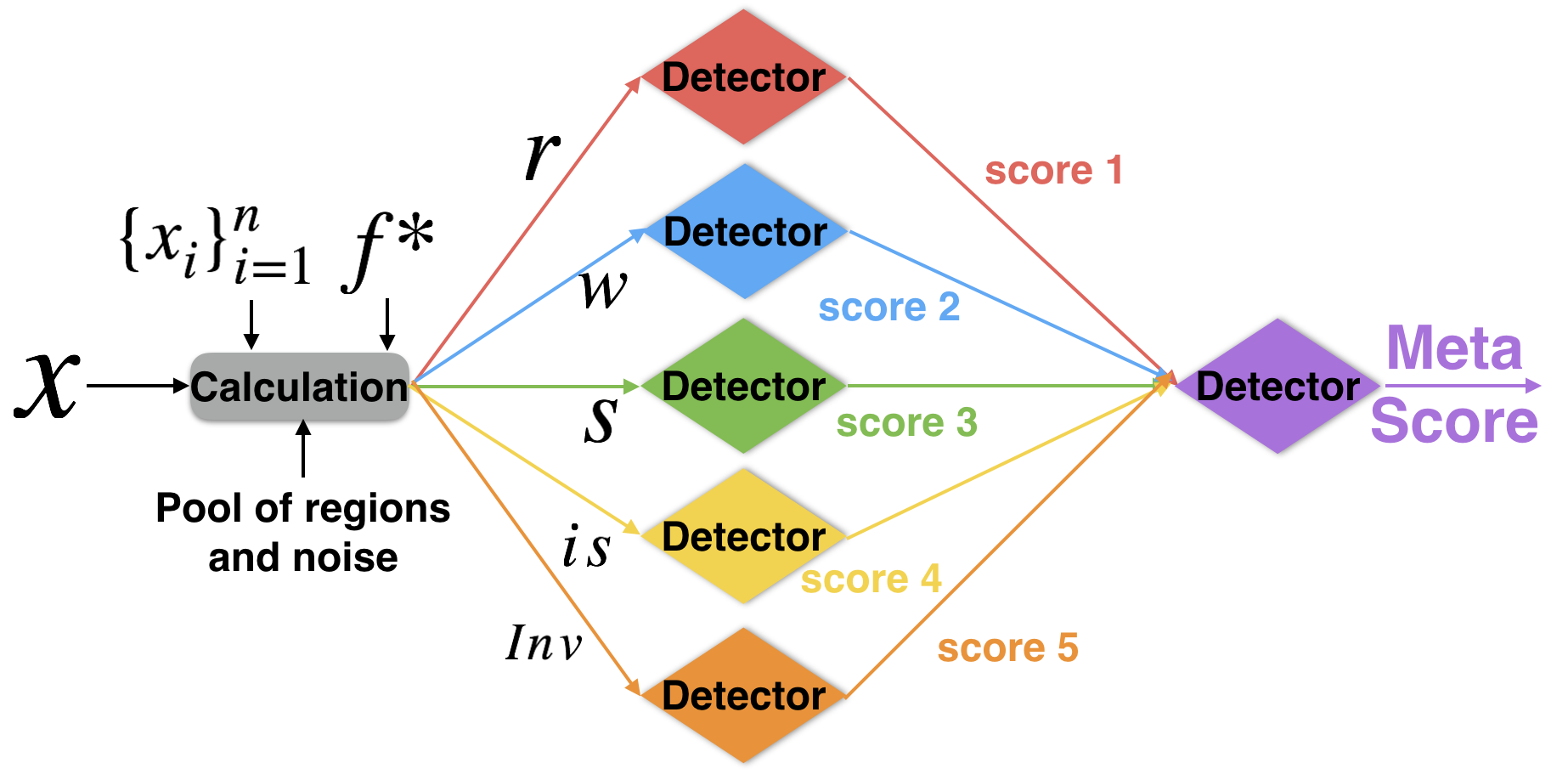}
    \caption{The novelty detection process: the five-metric values are first calculated for the input. Then each metric value is sent into a novelty detector that outputs a confidence score of the input not being a novelty. Lastly, five confidence scores are fed into a meta novelty detector to output a final confidence score.}
    \label{fig:process}
\end{figure}

The novelty detection process is shown in Fig.~\ref{fig:process}. For any input $z$, we first calculate the five-metric values using the validation data, the pool of central regions, and the noise variances via $MetricCal$ in Alg.~\ref{alg:calculating}. The metric values are  then sent into the corresponding novelty detectors, each of which will  output a confidence score of the input not being a novelty. A meta novelty detector takes all five confidence scores to output a final confidence score. By setting a user-defined threshold, poisoned inputs can be detected.   {We expect  each metric novelty detector to detect triggers in different ways.} The reason to use the meta novelty detector is that it fuses the confidence scores nonlinearly and is more accurate than a simple linear combination of the confidence scores.

In this paper, Local Outlier Factor (LOF) \cite{BKNS00} from scikit-learn \cite{scikit-learn} with default parameter settings is used as the novelty detector.   {The meta novelty detector is also a LOF with default parameter settings. However, our approach is applicable to any type of anomaly detector and does not necessarily require nearest-neighbor detectors. Indeed, we empirically observed that our approach is also accurate with one-class SVM \cite{CZH01} as the novelty detector. The number of LOF's parameters that need to be learned is small. Therefore, over-fitting is not likely to happen even if the available clean validation dataset is tiny. In contrast,   novelty detectors with a large parameter size (e.g., neural-network-based detectors) are likely to face over-fitting with small training datasets.} In this paper, the  clean validation dataset size is $30$. Therefore, the neural-network-based novelty detectors may be over-fitted and have low accuracy. Training the LOF is simple: after the training data is prepared, one  calls $model.fit$(training data)  for training. We used default values set by scikit-learn for all the training hyper-parameters. 

\begin{algorithm}
\caption{Detection Algorithm}  
 \label{alg:check} 
 \begin{algorithmic}
  \State {\bf Initialization:}
  \State Given validation dataset $\{x_i\}_{i=1}^n$ and network $f^*$
  \State {\bf Off-line training:}
  \State \#\#\# Calculating the Five-Metric Values for $\{x_i\}_{i=1}^n$ \#\#\#
      \State $\{r_l, w_l, s_l, is_l, Inv_l \}_{l=1}^n \leftarrow \{MetricCal(x_l, \{x_i\}_{i=1}^n) \}_{l=1}^n $
    \State \#\#\# Training the First Set of Novelty Detectors \#\#\#
    \State $\mathcal{N}_r.fit(\{r_l\}_{l=1}^{n})$, $\mathcal{N}_w.fit(\{w_l\}_{l=1}^{n})$, $\mathcal{N}_s.fit(\{s_l\}_{l=1}^{n})$
      \State $\mathcal{N}_{is}.fit(\{is_l\}_{l=1}^{n})$,  $\mathcal{N}_{Inv}.fit(\{Inv_l\}_{l=1}^{n})$
     \State \#\#\# Training the Meta Novelty Detector \#\#\#
     \State $\{Score_l\}_{l=1}^n$ $\leftarrow$ $\{(\mathcal{N}_r.score(r_l)$, $\mathcal{N}_w.score(w_l)$, $\mathcal{N}_s.score(s_l)$, $\mathcal{N}_{is}.score(is_l)$, $\mathcal{N}_{Inv}.score(Inv_l))\}_{l=1}^n$
     \State $\mathcal{N}_{meta}.fit(\{Score_l\}_{l=1}^n)$
     \State {\bf online detection:}
     \State Set a threshold $thres$
     \While {True}
     \State Given an input $z$;
     \State $r,w,s,is,Inv \leftarrow MetricCal(z, \{x_i\}_{i=1}^n)$
     \State $Score \leftarrow (\mathcal{N}_r.score(r),  \mathcal{N}_w.score(w),\mathcal{N}_s.score(s)$, \State $\mathcal{N}_{is}.score(is),\mathcal{N}_{Inv}.score(Inv))$
     \If {$\mathcal{N}_{meta}.score(Score)<thres$} 
        \State $z$ is poisoned, and the output should not be trusted
     \Else
     \State $z$ is clean, and  the output should be trusted
     \EndIf
     \EndWhile
 \end{algorithmic}
 \end{algorithm}
    
\subsection{The Detection Algorithm}
\label{sec:detection}
    Alg.~\ref{alg:check} describes how to train the novelty detectors and  use them to detect poisoned inputs. It first calculates the five-metric values for the clean validation samples $\{x_i\}_{i=1}^n$ using $MetricCal$ function. It then trains the five novelty detectors with the calculated metric values. The algorithm next feeds the calculated metric values to the trained novelty detectors to acquire the corresponding confidence scores. Finally, a meta novelty detector is trained with confidence scores. During online testing, the algorithm first calculates the metric values for a given input and then acquires the confidence scores by feeding the metric values into the first set of novelty detectors. It next feeds the scores into the meta novelty detector to get a meta score. If the meta score is lower than a user-defined threshold, the algorithm will consider $z$ poisoned. Otherwise, the input $z$ will be considered clean.

\section{Experimental Results}
\label{sec:exp}

\subsection{Datasets, Triggers, and Compared Methods}
\begin{table}
\centering
\caption{ {Clean and poisoned samples per class in training and testing datasets and clean validation dataset Size.}}
  {
\begin{tabular}{cccccccc}
\hline
&  & \multicolumn{2}{c}{Train / Class} & \multicolumn{2}{c}{Test / Class} & \multicolumn{2}{c}{Total Valid} \\
Dataset & Class & Clean & Poison  &  Clean & Poison & 1st & 2nd  \\
\hline
MNIST & 10 & 5500 & 825  & 1000 & 1000 & 30 & 100 \\
GTSRB &  43 & 820 & 123   &  294 & 294 & 30 & 100 \\
CIFAR-10 & 10 & 5000 & 750    & 500 & 500  & 30 & 100 \\
You. Face & 1283 & 81 & 12   &  10 & 10 & 30  & 100 \\ 
ImageNet & 200 & 500 & 50  &  10 & 10 & 30  & 100 \\
\hline
\end{tabular}
}
\label{table:size}
\end{table}

{\bf Clean Datasets and Network Architecture}: Our method is evaluated on various datasets, including MNIST \cite{MNIST}, GTSRB \cite{SSSI12}, CIFAR-10 \cite{KH09}, YouTube Face \cite{WHM11}, and a subset of ImageNet \cite{JWRLKL09}. The number of classes and the number of clean samples in the training and testing datasets for each class are shown in Table~\ref{table:size}.  For MNIST, the model was from \cite{GLDG19}. For GTSRB, the models were from \cite{WYSLVZZ19} and Pre-activation Resnet-18 \cite{HZRS16}. For CIFAR-10, the models were Network in Network \cite{LCY13} and Pre-activation Resnet-18. For YouTube Face, the network was from \cite{WYSLVZZ19}. For ImageNet, Resnet-18 was used. Since our method addresses a black-box scenario, any backdoored models (even models other than neural networks, such as SVM \cite{SVM} or random forest \cite{randomforest}) can be addressed. 

\begin{figure}
    \centering
    \hspace{-10mm}
    \includegraphics[width=\linewidth]{ 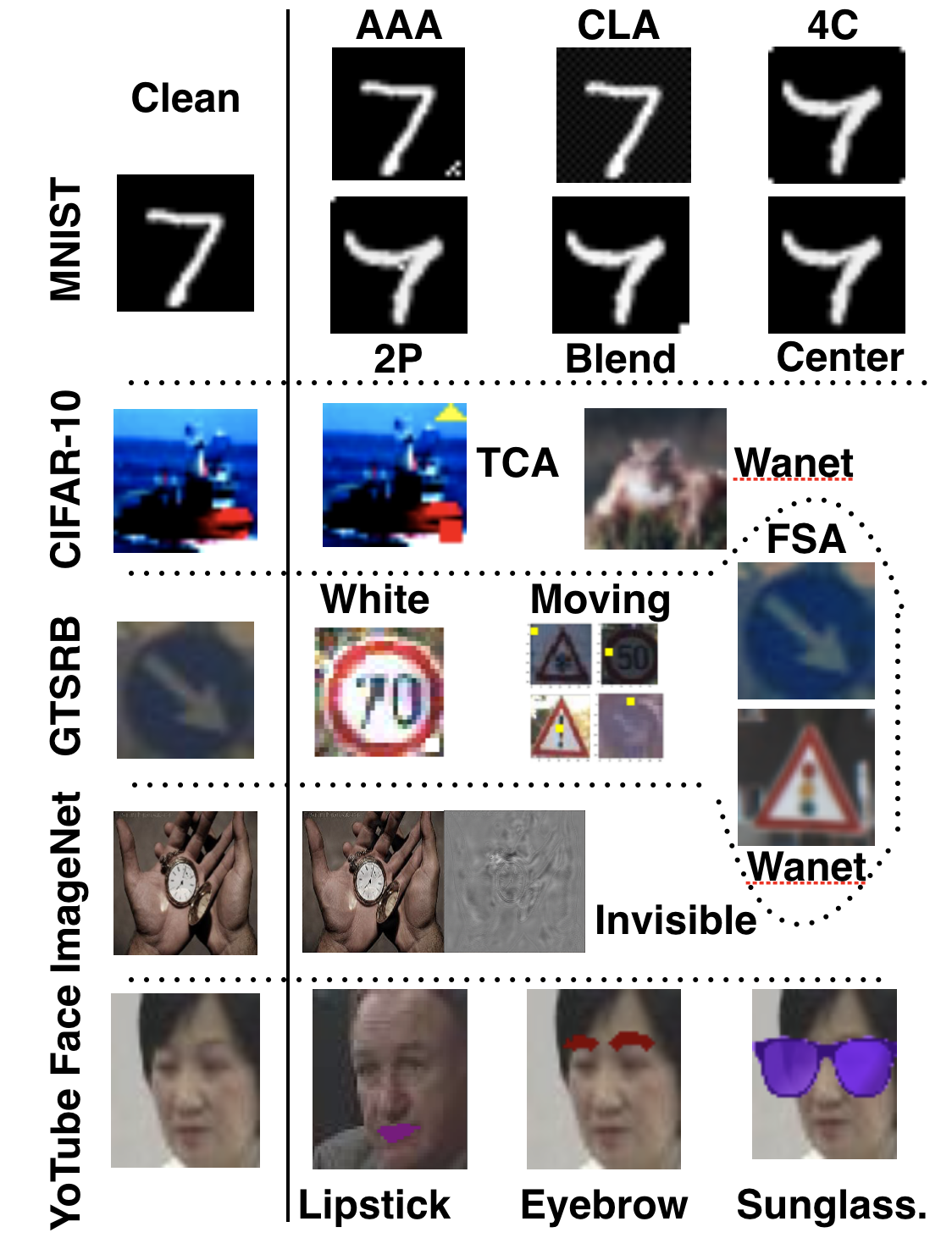}
    \caption{Clean and Poisoned Samples.}
    \label{fig:trigger}
\end{figure}

{\bf Triggers and Their Impacts}: The triggers for each dataset are shown in Fig.~\ref{fig:trigger}. In MNIST dataset, we considered all label attack (AAA) \cite{GLDG19}, clean label attack (CLA) \cite{ltkg20}, and blended (Ble.) \cite{CLLLS17}. Three additional triggers were created: the 4 corner trigger (4C) is four pixels at each corner; the 2 piece trigger (2P) is  2 pixels at the image's center and upper-left corner; the centered trigger (Cen.) is one pixel located in the image center.   {The desired impact of the triggers is to misguide the backdoored network to output the attacker-chosen label $l^*$ given in Table~\ref{table:optimal_asr}, where ``+1'' means $l^*= (l+1)$ mod 10.} In CIFAR-10 dataset, we considered combination attack (TCA) \cite{VLTKKKDG21} and Wanet (Wa.C) \cite{TA21} with the impact shown in Table~\ref{table:optimal_asr}. In GTSRB dataset, we considered a white box (Whi.) \cite{WYSLVZZ19}, a moving trigger (Mov.) \cite{FVKGK20,VLTKKKDG21, FVKGK22}, feature space attack (FSA) \cite{LLTMAZ19}, and Wanet (Wa.G) \cite{TA21}. In ImageNet dataset, we used an invisible (Invs.) trigger \cite{LLWLHL21}. In YouTube Face dataset, we used sunglasses (Sun.) \cite{CLLLS17}, lipstick (Lip.), and eyebrow (Eye.) \cite{FVKGK20,VLTKKKDG21, FVKGK22} as triggers.   {All the impacts (i.e., $l^*$) can be found in Table~\ref{table:optimal_asr}.
This paper later discusses the trigger patterns in more detail.}

  {
{\bf Details of Training Datasets}: We randomly injected the corresponding triggers into 15\% of the training clean samples for each class to create the training poisoned samples and changed the ground-truth label to the attacker-chosen label shown in Table~\ref{table:optimal_asr}. Table~\ref{table:size} shows the number of clean and poisoned samples per class in the training dataset. CIFAR-10, YouTube Face, and Sub-Imagenet are balanced datasets. Therefore, the numbers in Table~\ref{table:size} are the  numbers of samples per class in these three datasets. MNIST and GTSRB are not strictly balanced, but the numbers of samples per class for these two datasets are very close to the numbers shown in Table~\ref{table:size}. One can acquire more information on their distributions through the corresponding references. We followed the standard training process with CrossEntropy-based loss function and Adam optimizer to make the backdoored networks (badnet) have the classification accuracy (CA) and attack success rate (ASR) shown in Table~\ref{table:optimal_asr} ``Setup''.}

\begin{table*}
\begin{center} 
\caption{ {Badnet behavior and accuracy, ablation study in ROC (unless specified), and performance of using the adaptive threshold.}}
\label{table:optimal_asr}
   {
\begin{tabular}{cccc|ccccccc|cccc}
\hline
\multicolumn{4}{c|}{ {Setup}} & \multicolumn{7}{c|}{ {Comparison, Different Regions,  Replacing LOFs with OC-SVMs}} & \multicolumn{4}{c}{Adaptive Threshold (\%)} \\
\multicolumn{4}{c|}{ {Badnet Behavior and Accuracy (\%)}}  & \multicolumn{2}{c}{Ours} & \multicolumn{2}{c}{ {STRIP}} & \multicolumn{2}{c}{ {Add Extra Reg.}}  &  {OC-} &   \multicolumn{2}{c}{Our Method} &  \multicolumn{2}{c}{Applying}  \\
Trigger &  {$l^*$} & CA  & ASR & ROC &  {AUPR} &  {ROC} &  {AUPR} &  {$\mathcal{N}_w$} &  {Overall} &  {SVM} & TPR & FPR  & CA  & ASR  \\
\hline
AAA &  {+1} & 97.24 & 95.17 & 0.938 &  {0.974} &  {0.457} &  {0.601} &  {0.972} &  {0.972} &  {0.972} & 100 & 7.6 & 90.7 & 0 \\
CLA &  {0} & 89.1 & 100 & 0.934 &  {0.972} &  {0.914} &  {0.915} &  {0.781} &  {0.971} &  {0.969} & 100 & 8.31 & 87.09 & 0 \\
4C  &  {0} &  98.83 & 99.15 & 0.953 &  {0.973} &  {0.937} &  {0.935} &  {0.504} &  {0.973} &  {0.973} & 100 & 5.3 & 93.65 & 0 \\
2P &  {0}  & 99.17 & 94.71 & 0.965 &  {0.973} &  {0.959} &  {0.962} &  {0.524} &  {0.972} &  {0.962} &  94.1 & 1.35 & 97.93 & 1.1 \\
Ble. &  {0} & 98.45 & 99.55 & 0.974 &	 {0.971} &	 {0.910} &	 {0.908}	&  {0.973} &	 {0.970} &	 {0.972} & 91.35 & 2.45 & 96.32 & 8.54 \\
Cen. &  {0} & 98.96 & 92.21 & 0.989 &	 {0.974} &	 {0.782} &	 {0.778} &	 {0.972} &	 {0.972} &	 {0.972} & 92.6 & 1.05 & 98.04 & 3.01 \\
Whi. &  {33} & 96.55 & 97.4 & 0.969 &	 {0.972} &	 {0.932} &	 {0.942}	&  {0.972} &	 {0.971} &	 {0.971} & 98.1 & 2.95 & 94.8 & 0.05 \\
Mov. &  {0} & 95.15 & 99.88 & 0.988 &	 {0.973} &	 {0.845} &	 {0.854}	&  {0.707} &	 {0.973} &	 {0.972} & 99.75 & 1.01 & 94.62 & 0 \\
FSA &  {35} & 95.03 & 90.2 & 0.930 &	 {0.935} &	 {0.402} &	 {0.423}	&  {0.777} &	 {0.928} &	 {0.940} & 90.2 & 4.9 & 91.34 & 0 \\
TCA &  {7} & 88.9 & 99.7 & 0.965 &	 {0.973} &	 {0.967} &	 {0.970} &	 {0.485} &	 {0.973} &	 {0.970} & 95.45 & 1.85 & 87.65 & 4.27 \\
Sun. &  {0} & 97.85 & 100 & 0.986 &	 {0.934} &	 {0.918} &	 {0.925} &	 {0.559} &	 {0.970} &	 {0.949}  & 91.55 & 0.95 & 97.3 & 8.45 \\
Lip. &  {0} & 97.3 & 91.39 & 0.962 &	 {0.957} &	 {0.881} &	 {0.898} &	 {0.732} &	 {0.947} &	 {0.946}  & 88.7 & 0.83 & 96.81 & 2.85 \\
Invs. &  {0} & 78.41 & 99.94 & 0.984 &	 {0.972} &	 {0.875} &	 {0.890}	&  {0.835} &	 {0.970} &	 {0.970} & 97.61 & 2.53 & 77.8 & 2.33 \\
Wa.C &  {0} & 93.9 & 99.45 & 0.924 &	 {0.937} &	 {0.510} &	 {0.507} &	 {0.422} &	 {0.935} &	 {0.928} & 87.14 & 11.2 & 85.13 & 12.36 \\
Wa.G &  {0} & 99.05 & 99.45 & 0.992 &	 {0.972} &	 {0.424} &	 {0.419}	&  {0.482} &	 {0.970} &	 {0.969}  & 97.65 & 1.23 & 98.17 & 1.8\\
\hline 
\end{tabular}
}
\end{center}
\end{table*}

\begin{figure}
    \centering
    \includegraphics[width=\linewidth]{ 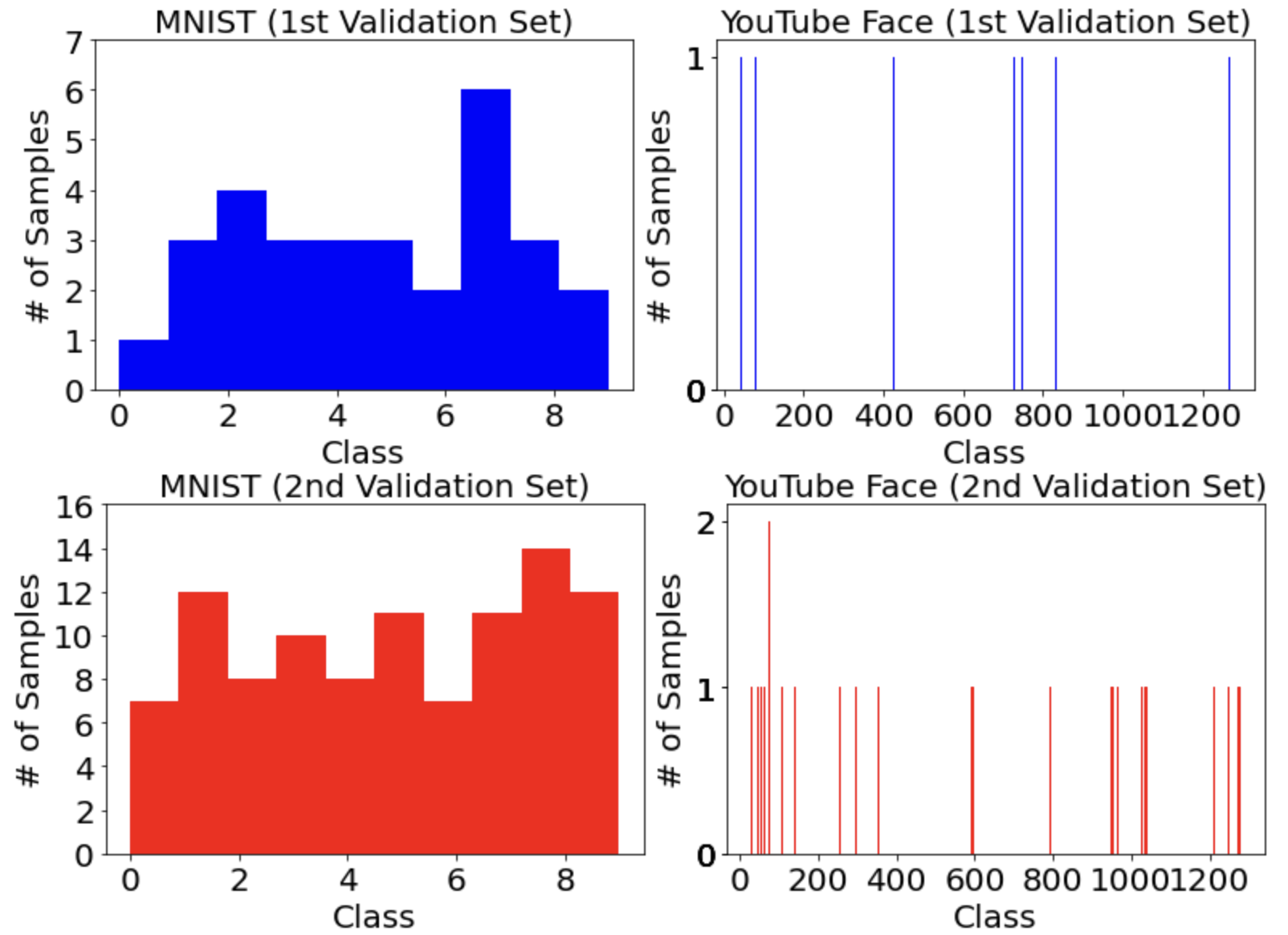}
    \caption{ {Histograms of clean validation datasets.  {Validation samples are randomly selected from the clean datasets.} Blue: $n=30$. Red: $n=100$. }}
    \label{fig:distributions}
\end{figure}

  {
{\bf Details of Validation and Testing Datasets}: 
Our work considers two small validation datasets with sizes $n=30$ and $100$ for all cases as shown in Table~\ref{table:size}. The first clean validation datasets were used for training all the novelty detectors.  { In this work, we randomly select clean samples to form the validation datasets so that the clean validation datasets have the same underlying data distributions as the clean testing datasets. We do not require the clean validation samples to be 100\% correctly classified by the model. }  The number of samples per class for MNIST and YouTube Face is shown in the first row of Fig.~\ref{fig:distributions}.  CIFAR-10 and MNIST have similar numbers of clean validation samples per class because they both have 10 classes, whereas the numbers of clean validation samples per class for GTSRB and sub-ImageNet are similar to YouTube Face because they all have a large number of classes. The second validation datasets were utilized to determine a proper threshold for our approach and have the same underlying distributions as the clean testing datasets as well. The second row in Fig.~\ref{fig:distributions} shows the number of samples per class in the second clean validation datasets of MNIST and YouTube Face.  Table~\ref{table:size} also shows the testing datasets that are used to evaluate our approach and other compared methods. The ratio of poisoned samples to clean samples is 1 (i.e., the testing datasets are balanced). Specifically, we generated the poisoned testing samples by injecting triggers into their corresponding clean versions. It is worth noticing that methods evaluated on imbalanced binary classification datasets may have high inference accuracy but poor run-time performance.
}

{\bf Compared Methods}: We selected one reverse-engineering-based approach (i.e., Neural Cleanse \cite{WYSLVZZ19}), one out-of-distribution detection (i.e.,  Mahalanobis-distance-based novelty detection (MD) \cite{LLLS18}), one retraining-based approach (i.e., Kwon's method \cite{K20}), and STRIP \cite{GXWCRN19}.   {Neural Cleanse directly modifies the parameters of the original backdoored network to cap the maximum neuron values for the reverse-engineered triggers. Kwon's method trains a new clean network on some relabeled poisoned samples. STRIP utilizes an entropy-based confidence score and ``blend'' technique for backdoor detection. MD is a feature-based anomaly detector with a Gaussian-based confidence score. MSP \cite{DK17} and GEM \cite{ML22} are two additional feature-based anomaly detectors whose performance was found to be close to MD. Therefore, for brevity, we only present the results with MD.} The compared methods are representative in their types of defenses. 

{\bf Fitting the Novelty Detectors}: We used LOFs for the metric and meta novelty detectors. However, other types of novelty detectors are also allowed, such as one-class SVM. The training process is shown in Alg.~\ref{alg:check}. Only the clean validation datasets  with size $n=30$ are available. Neither poisoned samples nor information about triggers were used.  All the training hyper-parameters were set to default values provided by scikit-learn \cite{scikit-learn}.

\subsection{Ablation Study for the Five Metrics}
\begin{figure*}
    \centering
    \includegraphics[width=\textwidth]{ 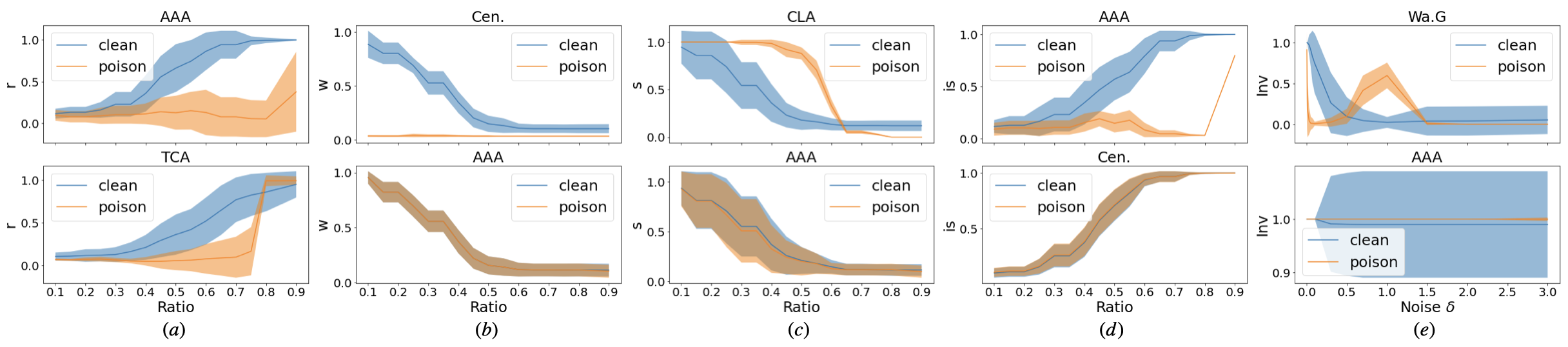}
    \caption{  {The five metrics values of clean and poisoned samples for different backdoor attack cases. The plots show the average of multiple runs utilizing multiple samples in each run.  The first row shows the cases where clean and poisoned samples are distinguishable with respect to each metric value. The second row shows the cases where clean and poisoned samples are not distinguishable with respect to each metric value. (a): the visualization of \eqref{robustness} for  $r$. (b): the visualization of \eqref{weakness} for  $w$. (c): the visualization of \eqref{sensitivity} for  $s$. (d): the visualization of \eqref{inverse} for  $is$. (e) the visualization of \eqref{noise} for  $Inv$.} }
    \label{fig:illustration}
\end{figure*}

\begin{figure*}
    \centering
    \includegraphics[width=\textwidth]{ 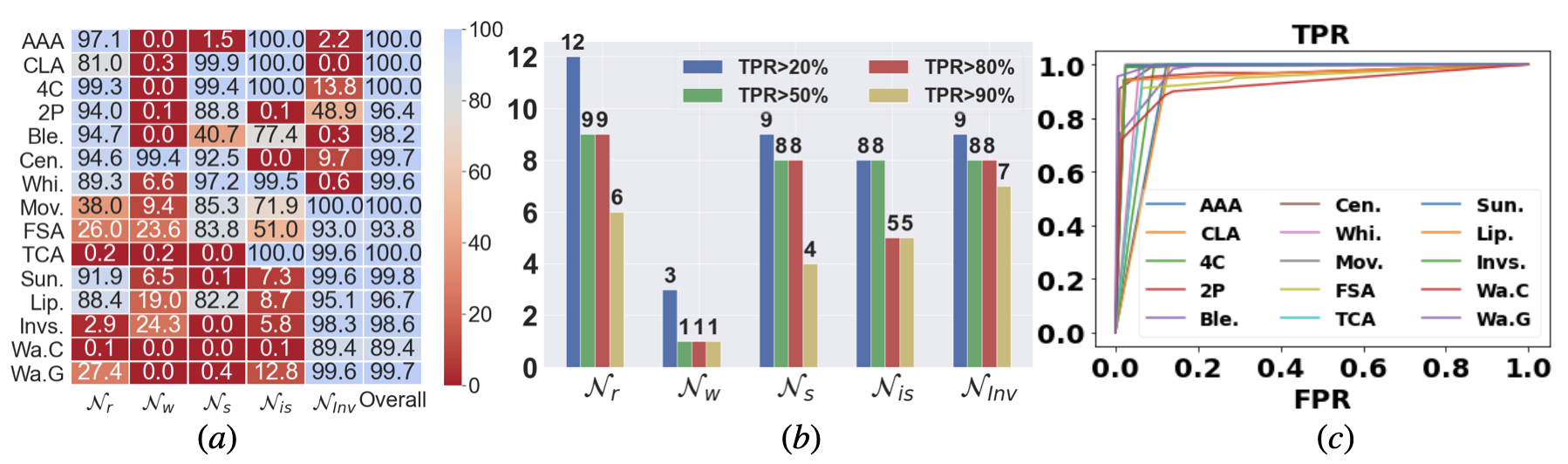}
    \caption{(a): TPR of the five metric novelty detectors.  Overall: $or$ operation of the five novelty detectors. Higher numbers indicate better performance. (b): The total number of effective cases for each metric. (c): ROC curve for the meta novelty detector. }
    \label{fig:heatmap}
\end{figure*}

\begin{figure}[ht]
    \centering
    \includegraphics[width=0.9\linewidth]{ 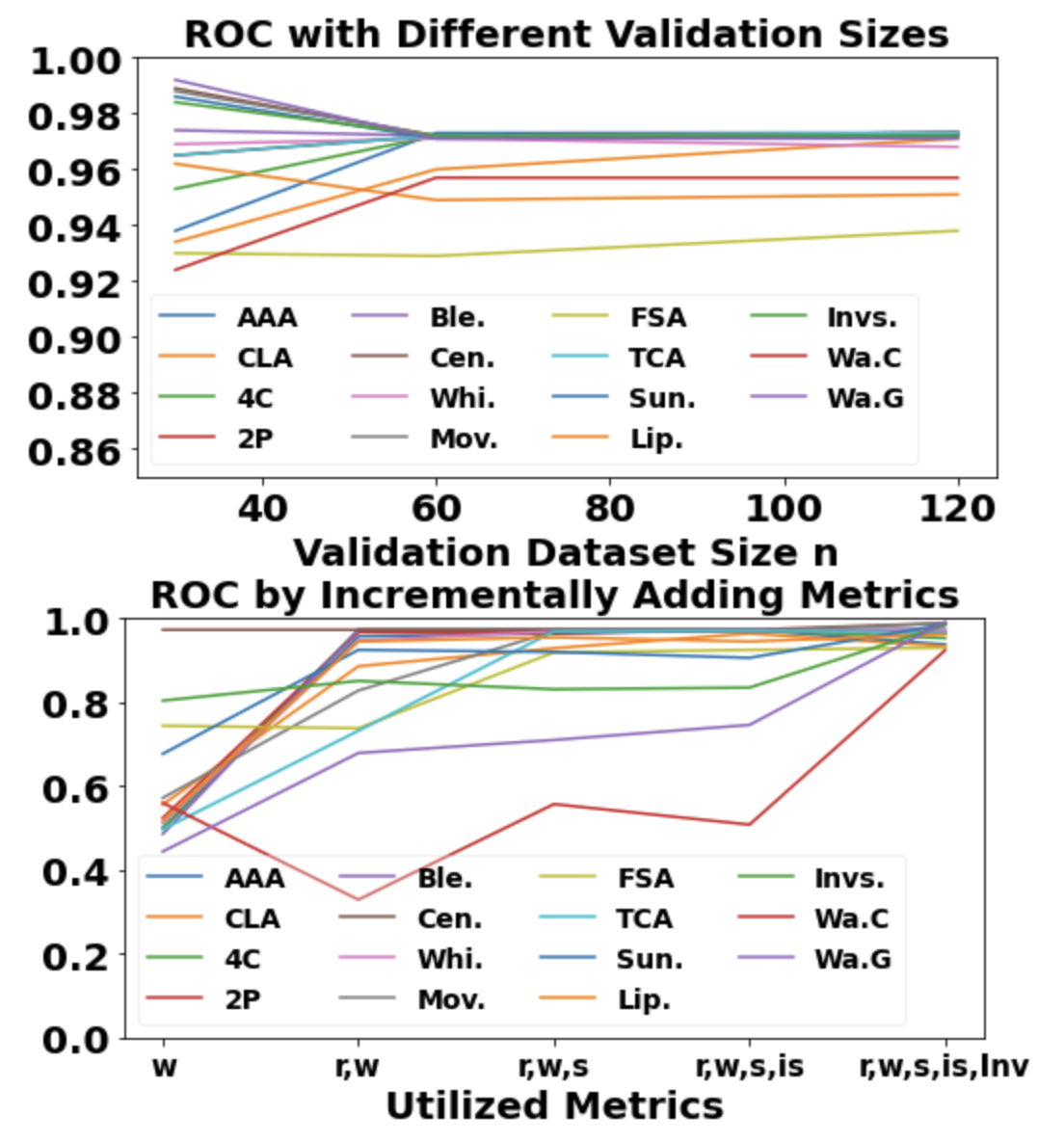}
    \caption{ {AUROC of the meta novelty detector with different clean validation dataset sizes $n$ (up) and incrementally adding metrics  (bottom).}}
    \label{fig:ablation}
\end{figure}

  {
The ablation study shows the efficacy of each metric detector, the reasoning behind the validation dataset size and central regions, and improvement by incrementally adding metrics. The results are shown in Table~\ref{table:optimal_asr} and Figs.~\ref{fig:illustration}-\ref{fig:ablation}.
}

{\bf How Each Metric Works}:  { To show the efficacy of the introduced five metrics given by \eqref{robustness}, \eqref{weakness}, \eqref{sensitivity}, \eqref{inverse}, and \eqref{noise}, we have plotted the five metrics values of clean and poisoned samples with respect to different aspect ratios (small ratios represent small central regions) and noise variances (small variances represent small noise perturbations) for several backdoor attack cases. The results are shown in Fig.~\ref{fig:illustration}. The first row shows the cases where clean and poisoned samples are distinguishable with respect to each metric value. Based on this observation, it is important to use several central regions with different sizes together to increase the detection probability of poisoned samples. For example, in the case AAA, clean and poisoned samples have similar $r$ and $is$ values for small central regions. However, they have distinguishable $r$ and $is$ values for large central regions. As for the case Cen., clean and poisoned samples are distinguishable in terms of $w$ for small and medium central regions. Clean and poisoned samples in CLA can be distinguished in terms of $s$ using medium central regions. Clean and poisoned samples in Wa.G can be distinguished by the metric $Inv$ using small and medium noise perturbations. The second row in Fig.~\ref{fig:illustration} shows the cases where clean and poisoned samples are not distinguishable with respect to each metric value for all central regions and noise variances. Therefore, it is important to use all the five metrics together to increase the detection probability of poisoned samples. Details are given in the following ablation study.}

{\bf Contribution of Each Metric}: We calculated the five-metric values of the  poisoned samples in the testing data for each backdoor attack case and input them into the five novelty detectors. After receiving the confidence scores, we calculated the TPR of each novelty detector using the default threshold value provided by scikit-learn (i.e., 0). Fig.~\ref{fig:heatmap}(a) shows the TPR for each backdoor attack case. The ``Overall'' column is the final TPR of $or$ operation of the five novelty detectors. It is seen that each metric contributes to detecting poisoned inputs for some cases. Combining the five metrics can achieve a high TPR of over 90\% for most cases except Wanet CIFAR-10 (Wa.C) whose TPR is 89.4\%. 

Fig.~\ref{fig:heatmap}(b) shows the number of cases in which each metric contributes to detecting poisoned samples over the total 15 cases. $\mathcal{N}_r$ has a TPR of more than 20\% in over 80\% the cases. $\mathcal{N}_w$ helps increase the ``Overall'' TPR for some cases. For example, without using $\mathcal{N}_w$, the TPR reduces by 3\% in both ``Cen.'' and ``Lip.'' cases.    {We also observed that $\mathcal{N}_w$ improved its own performance in some triggers if additional regions are utilized as shown in  Table~\ref{table:optimal_asr} ``$\mathcal{N}_w$''.}  For more than half the cases, $\mathcal{N}_s$ and $\mathcal{N}_{is}$ have TPR of over 50\%.    {Besides the non-pattern-based triggers, $\mathcal{N}_{Inv}$ is also effective on some pattern-based triggers since the generated noise breaks the impact of some pattern-based triggers, leading to an abrupt change in $Inv$ and the detection by $\mathcal{N}_{Inv}$.}

The FPR of using the $or$ operation in ``Overall'' column is less than 30\% for most cases. However, our meta novelty detector can reach a better trade-off between TPR and FPR than a simple $or$ operation of the five novelty detectors.

{\bf Performance of the Meta Novelty Detector}: Choosing different threshold values $thres$ will lead to different TPR and FPR. We, therefore, draw the ROC curves for the meta novelty detector shown in Fig.~\ref{fig:heatmap}(c) with the AUROC shown in Table~\ref{table:optimal_asr}. The meta novelty detector can reach an AUROC over 0.9 for all the triggers. Therefore, for a proper threshold value, the meta novelty detector will have a high TPR and a low FPR.   {Although the testing datasets are balanced with an equal number of clean and poisoned samples, we still show the AUPR to investigate the performance of our approach in finding the poisoned samples. From Table~\ref{table:optimal_asr}, our approach also has high AUPRs. Compared with the baseline method STRIP, our approach shows consistently high performance in all the cases, whereas STRIP performs poorly in several cases, such as AAA, Wa.C, and Wa.G. We also applied one-class SVM (OC-SVM) as the metric detectors and the meta detector and show the result in Table~\ref{table:optimal_asr}. Based on the result, our approach is applicable to any type of anomaly detector and does not necessarily require LOFs.}

{\bf Regions, Sizes, and Incremental Improvements}: We evaluated our approach by using additional regions. Specifically, the coordinates of the additional regions are   $m_1 = (1-k)L$, $n_1 = (1-k)W$, $m_2 = L$, and $n_2= W$, where $k$ is the aspect ratio from the same ratio pool in Alg.~\ref{alg:calculating}. The results are shown in Table~\ref{table:optimal_asr} ``Overall''. The AUROC of using extra regions is close to using only the central regions. Therefore, to minimize the computation cost, we choose to use only the central regions.  The first plot in Fig.~\ref{fig:ablation} shows the performance of our approach with different validation sizes. Using more validation samples does not significantly improve our approach but adds memory and computation complexity. Therefore, we consider $n=30$ to be optimal. The second plot in Fig.~\ref{fig:ablation} shows the improvement of our approach by incrementally adding metrics. The AUROC is over 0.9 by using only two metrics ($w$ and $r$) in eight cases. With three metrics ($w$, $r$, and $is$), the AUROC is over 0.9 in twelve cases. Therefore, our approach requires metrics less than five to be accurate in some cases. Our approach in the remaining three cases requires all five metrics to achieve over 0.9 AUROC (using only $r$, $w$, $s$, and $is$, the AUROC of our approach is 0.835 for Invs., 0.508 for Wa.C, and 0.746 for Wa.G).  { To further highlight the contribution of each metric, we have evaluated our approach on all the cases by using only four metrics. The Sun. case mostly highlights the contribution of each metric. Therefore, this paper  mainly discusses the Sun. case  for brevity.  The AUROC is 0.571 for using only $w$, $s$, $is$, $Inv$, 0.937 for using only $r$, $s$, $is$, $Inv$, 0.895 for using only $r$, $w$, $is$, $Inv$,  0.905 for using only $r$, $w$, $s$, $Inv$, and 0.906 for using only $r$, $w$, $s$, $is$. The AUROC of using all the five metrics is 0.986. According to these numbers, the metric $r$ contributes to detecting poisoned samples most for the Sun. case. However, the other four metrics also greatly fine-tune the performance of our approach by increasing the AUROC by roughly  5\% $\sim$ 8\%.  Therefore, all the five metrics are needed to maximally capture poisoned samples. }

\subsection{Adaptive Search for  Threshold}

Recall that our approach requires an appropriate threshold
value is required to have a high TPR and a low FPR.  To find such a threshold, we propose an adaptive-search solution. After the meta novelty detector is trained, one collects all the meta scores of the clean validation data (i.e., $\mathcal{N}_{meta}.score(\{Score_l\}_{l=1}^n$)). One finds the mean $\mu$ and standard deviation $\sigma$ of these meta scores. The threshold value can be set to $thres = \mu - h*\sigma$, where $h$ is a coefficient. With the second validation dataset available as shown in Table~\ref{table:size}, the user can vary $h$ and observe the FPR by testing the meta novelty detector on this validation dataset. Then, the user can choose the desired $h$ according to the corresponding FPR. If the second validation dataset is not available, the user can set $h$ to be any reasonable number. 

With this adaptive search, our detection algorithm can reach the FPR of less than 5\% and TPR of more than 90\% for most cases, as shown in Table~\ref{table:optimal_asr}. The CA and ASR of the backdoored network before and after applying our method are also shown in Table~\ref{table:optimal_asr}. It is seen that our approach reduces the ASR by identifying and discarding potential poisoned inputs and maintains a reasonable CA. Even though our approach has a relatively low TPR (88.78\%) for the trigger Lipsticks, the ASR for the Lipsticks is low (2.85\%). This is because there exist null poisoned samples  that fail to make the backdoored network output the attacker-chosen label. Since our approach is based on detecting differences in behaviors of the network when presented with triggers vs. benign features, the null poisoned samples may be considered clean by our approach and bypass the detection. However, failing to detect null poisoned samples does not increase the ASR.  {Note that the adaptive-search method can find  better thresholds if a larger second validation dataset is  used. For example, in Sun., the ASR can be further reduced to 0.9\% with the CA being 95\%. }

\subsection{Data-Efficiency and Comparisons}

\begin{table*}
\centering
\caption{ Original badnet performance and comparisons with compared methods  on selected triggers and muti-trigger attacks.}
\begin{tabular}{ccccccccccccc}
\hline
\noalign{\smallskip}
 & \multicolumn{2}{c}{Bad Net.}  & \multicolumn{2}{c}{Ours} & \multicolumn{2}{c}{Neural Cleanse \cite{WYSLVZZ19}} & \multicolumn{2}{c}{MD \cite{LLLS18}} & \multicolumn{2}{c}{Kwon's \cite{K20}} & \multicolumn{2}{c}{STRIP \cite{GXWCRN19}} \\
Trigger & CA & ASR  & CA  & ASR & CA & ASR & CA & ASR & CA & ASR & CA & ASR \\
\hline
AAA  & 97.24 & 95.17 & 90.7 & 0 & 97.25 & 95.06 & 0 & 0 & 52.49 & 5.2  & 96.71 & 94.22 \\
CLA & 89.1 & 100  & 87.09 & 0 & 86.65 & 100 & \multicolumn{2}{c}{Fail} & 54.88 & 5.69 & 87.44 & 55.37 \\
\hline 
Whi. & 96.55 & 97.4  &  94.8 & 0.05 & 96.55 & 97.44 & \multicolumn{2}{c}{Fail} & 25.45 & 0 &  96.03 & 97.13 \\
Mov. & 95.15 & 99.88 &  94.62 & 0 & 95.2 & 99.88 &  \multicolumn{2}{c}{Fail} & 20.13 & 0 & 95.41 & 100 \\
FSA & 95.03 & 90.2 &  91.34 & 0 &  94.80 & 89.43 & \multicolumn{2}{c}{Fail} & 18.93 & 0 & 94.11 & 90.3 \\
\hline
TCA & 88.9 & 99.7 &  87.65 &  4.27 & 87.92 & 99.68 & 70.72 & 86.58 & 15.3 & 0 &  81.92 & 0.25 \\
\hline
Sun.  & 97.85 & 100 & 97.3 & 8.45 & 97.59 &  99.90 & \multicolumn{2}{c}{Fail }  & 1.1 & 0 & 90.74 & 7.02  \\
Lip. & 97.3 & 91.39 &  96.81 & 2.85 & 96.81 & 91.41 & \multicolumn{2}{c}{Fail} & 1.1 & 0 & 90.93 & 6.91 \\
\hdashline
\multirow{3}{*}{\shortstack[c]{Sun., Lip., Eye. \\ $l^*$}} & \multirow{3}{*}{95.90}  & 92.2 & \multirow{3}{*}{92.15} & 0 & \multirow{3}{*}{N/A} & \multirow{3}{*}{N/A} & \multirow{3}{*}{N/A} & \multirow{3}{*}{N/A} & \multirow{3}{*}{N/A} & \multirow{3}{*}{N/A} & \multirow{3}{*}{91}  & 3.7 \\
&  & 92.2 &  & 0.5 &  &  &  &  &  &  &  & 41.7 \\
&  & 100 &  & 0 &  &  &  &  &  &  &  & 0 \\
\hdashline
\multirow{3}{*}{\shortstack[c]{Sun., Lip., Eye. \\ $l_1^*,l_2^*,l_3^*$}} & \multirow{3}{*}{95.94}  & 91.5 & \multirow{3}{*}{92.96} & 0.3 & \multirow{3}{*}{N/A} & \multirow{3}{*}{N/A} & \multirow{3}{*}{N/A} & \multirow{3}{*}{N/A} & \multirow{3}{*}{N/A} & \multirow{3}{*}{N/A} & \multirow{3}{*}{90.5}  & 3.8 \\
&  & 91.3 &  & 0 &  &  &  &  &  &  &  & 32.3 \\
&  & 100 &  & 0 &  &  &  &  &  &  &  & 0 \\
\hline
Invs. & 78.41 & 99.94 & 77.8  & 2.33 & N/A & N/A & 0 & 0 & N/A & N/A & 74.01 & 24.13  \\
\hline
\end{tabular}
\label{table:results} 
\end{table*}

We selected several types of triggers  and compared our approach with other methods. We trained the baseline models with the two tiny validation datasets (i.e., a total of 130 samples) for a fair comparison. The hyper-parameters for training the compared methods were set based on the original papers and the codes provided by the authors.   {We set the thresholds for STRIP and MD so that  5\% of the clean validation samples are considered poisoned.} The results are summarized in Table~\ref{table:results}.

{\bf Naive Triggers}: The  naive triggers (i.e., CLA and Whi.) are  simple patterns associated with an attacker-chosen label. Our method  reduces the ASR to a low value (maximum is 0.05\%) while maintaining a reasonable CA. The other methods either do not work or cannot achieve comparable CA or ASR.

{\bf Functional Complex Triggers}: According to the $l^*$ in Table~\ref{table:optimal_asr}, the attack AAA depends on the trigger pattern and the image's benign features. Additionally, AAA is a one-to-all attack with attacker-chosen label $l^* = l+1$ mod 10.  The trigger of Mov. is randomly attached to the image. The trigger for TCA is the combination of two different shapes, and the network will output the attacker-chosen label only when both shapes exist. If only one shape appears in the image, the network  behaves normally.  The compared methods cannot both detect poisoned inputs and maintain high CA.

{\bf Real-World Meaning Triggers}: The attacker   uses some real-world objects as the triggers (i.e., Sun. and Lip.), and the trigger size can be large.  Neural Cleanse cannot reverse-engineer large-sized triggers and thus has low accuracy. This is verified in Table~\ref{table:results}. STRIP is still valid in this case, but our method shows a higher CA. 

{\bf Filter Trigger and Invisible Sample-Specific Trigger}: The trigger for FSA is a Gotham filter. Inputs that pass through this filter become poisoned. This trigger essentially changes the entire input image. However, our method shows its efficacy while all other methods fail. Trigger Invs. is an invisible sample-specific trigger. The attacker extracts  content information and generates a hidden pattern for each image (the last picture in the ``ImageNet'' row in Fig.~\ref{fig:trigger}). By injecting the hidden pattern, the poisoned input looks identical to the clean input to human eyes (the middle picture in the ``ImageNet'' row), and  the network will output the attacker-chosen label. Our method can detect this advanced attack as well.

{\bf Information Comparison}: Table~\ref{table:info} shows the information needed for each method to detect or defend against backdoor attacks. For reverse-engineering-based approaches such as Neural Cleanse, they need access to the network parameters. However, it may not always be possible since such information could be proprietary. Feature-based statistical detection tools, such as MD, require some hidden layer outputs of the network. Compared to reverse-engineering-based approaches, they require less information about the network. Nevertheless, they may not be viable if the network is proprietary. The retraining-based approaches, such as Kwon's, need the architecture of the network. However, to achieve  high performance, they need a large amount of clean data, which may not be possible. Lastly, STRIP is a statistical detection tool that  requires only the logits layer (i.e., the layer before the softmax function) output. However, if a neural network is entirely black-box,  STRIP is also  inapplicable. In contrast, our  method can operate in completely black-box scenarios (black-box-efficient) and with smaller amounts of clean validation data (data-efficient).

\begin{table}[ht]
\begin{center} 
\caption{Required information    for each detection method.   }
\label{table:info}
\begin{tabular}{ccccc}
\hline
\noalign{\smallskip}
Approach & Hidden Layer Output &  Logits & Output Label \\
\hline
Ours & No &  No & Required \\
Neural Cleanse &  Required &  Required & Required\\
MD & Required & No & No\\
Kwon's & No &  No & Required \\ 
STRIP & No &  Required & No \\
\hline
\end{tabular}
\end{center}
\end{table}

{\bf Reasons for Low Accuracy on the Compared Methods}: The most important reason for their low accuracy is that the available clean validation dataset is small. Neural Cleanse and Kwon's method  need to fine-tune a neural network model. The two tiny validation datasets are not enough to fine-tune neural network models to have high accuracy. MD and STRIP do not use neural networks. However, MD trains a novelty detector with hidden layer outputs of the validation samples, which are high-dimensional vectors (e.g., a 100-dimensional vector). The small number of validation samples and the high dimension of the hidden layer outputs make the trained novelty detector have low accuracy. STRIP uses the ``blend'' function to create synthetic images. However, there are many triggers whose functionality can be broken by the ``blend'' function. Therefore, STRIP becomes ineffective on those triggers. For example, STRIP is not accurate for one-to-all cases (i.e., AAA). Our method utilizes multiple regions to extract image contents to avoid breaking the triggers' functionality.

\subsection{Benign Models and Multiple Triggers}
Although our method does not detect if the model is backdoored and instead detects potentially poisoned inputs, other works (e.g., Neural Cleanse) can be used for checking if a model is backdoored. If the model is indeed backdoored, our method still applies to find poisoned inputs during online operation while using the model. If the model is clean but misidentified as backdoored, our method can retain a reasonable CA value. For example, we trained a benign model with MNIST dataset, which has 98.95\% CA. After applying our method, the CA becomes 95.86\%, which is still reasonable. Therefore, it is safe to use our method even when there are no backdoor attacks. We also considered multi-trigger-single-target attack (MTSTA) and multi-trigger-multi-target attack (MTMTA). In MTSTA, there are three triggers associated with a single attacker-chosen label. In MTMTA, the three triggers are associated with three different attacker-chosen labels. As shown in Table~\ref{table:results}, our method works for both cases for all triggers. However, STRIP fails to detect some  triggers.

\section{Adaptive Attacks and Future Works}
The proposed metrics help understand the backdoored network behavior.  To bypass our detection, a backdoor attack needs to satisfy several conditions. Its trigger should be non-pattern-based since our approach uses four metrics to detect pattern-based triggers and attains high accuracy.  Additionally, the robustness of the trigger against noise perturbation should be close to benign features so that the metric values $Inv$ for clean and poisoned samples are similar. The attacker can attempt to design an adaptive attack by using the five metrics  in the training loss function. While we have not so far been able to devise a straightforward way to construct such an adaptive backdoor attack, it appears that the Wanet in CIFAR-10 case provides some clue in this direction since it provided a relatively low CA and high ASR compared to other attacks (although we did reasonably well in this case as well).   

 {From the ablation studies, it appears that one metric $Inv$ is dominant for detecting non-pattern-based triggers although other metrics have some contribution. One potential direction for future work is to  add new  metrics for non-pattern-based triggers. For example,  the denoising technique may also  contribute to detecting non-pattern-based triggers.   Another potential direction is to build new deep novelty detectors based on the existing ones to capture poisoned samples more accurately. The deep novelty detectors have shown promising performance for many applications. However, there are several factors that limit the existing deep novelty detectors to be utilized on detecting poisoned samples under the considered scenario. One factor is that training the deep novelty detectors requires sufficiently a large amount of data. Deep networks, such as LeNet, have thousands of parameters for even a single layer. However, the available clean samples considered in this work are scarce (i.e., $n\le 30$). Therefore, the overfitting is likely to happen when using the existing deep novelty detectors for backdoor detection with limited data. Another factor is that the existing deep novelty detectors' accuracy still requires improvement. Therefore, building new deep novelty detectors that  are less demanding for data and more accurate for backdoor detection can be fruitful.}

\section{Conclusion}
\label{sec:con}

The behavioral differences between triggers and benign features are illustrated and utilized to detect backdoored networks. Five metrics are proposed to measure the behavior of the network for a given input. A novelty detection process is proposed to detect poisoned inputs by taking as input the five-metric values. The method is black-box-efficient and data-efficient.  The ablation study for the five metrics, the efficacy of our approach, and the comparison with other  methods are shown on various types of backdoor attacks. Potential adaptive attacks and prospective works are also discussed.

{\small
\bibliographystyle{ieee_fullname}
\bibliography{egbib}
}

\begin{IEEEbiography}[{\includegraphics[width=1in,height=1.25in,clip,keepaspectratio]{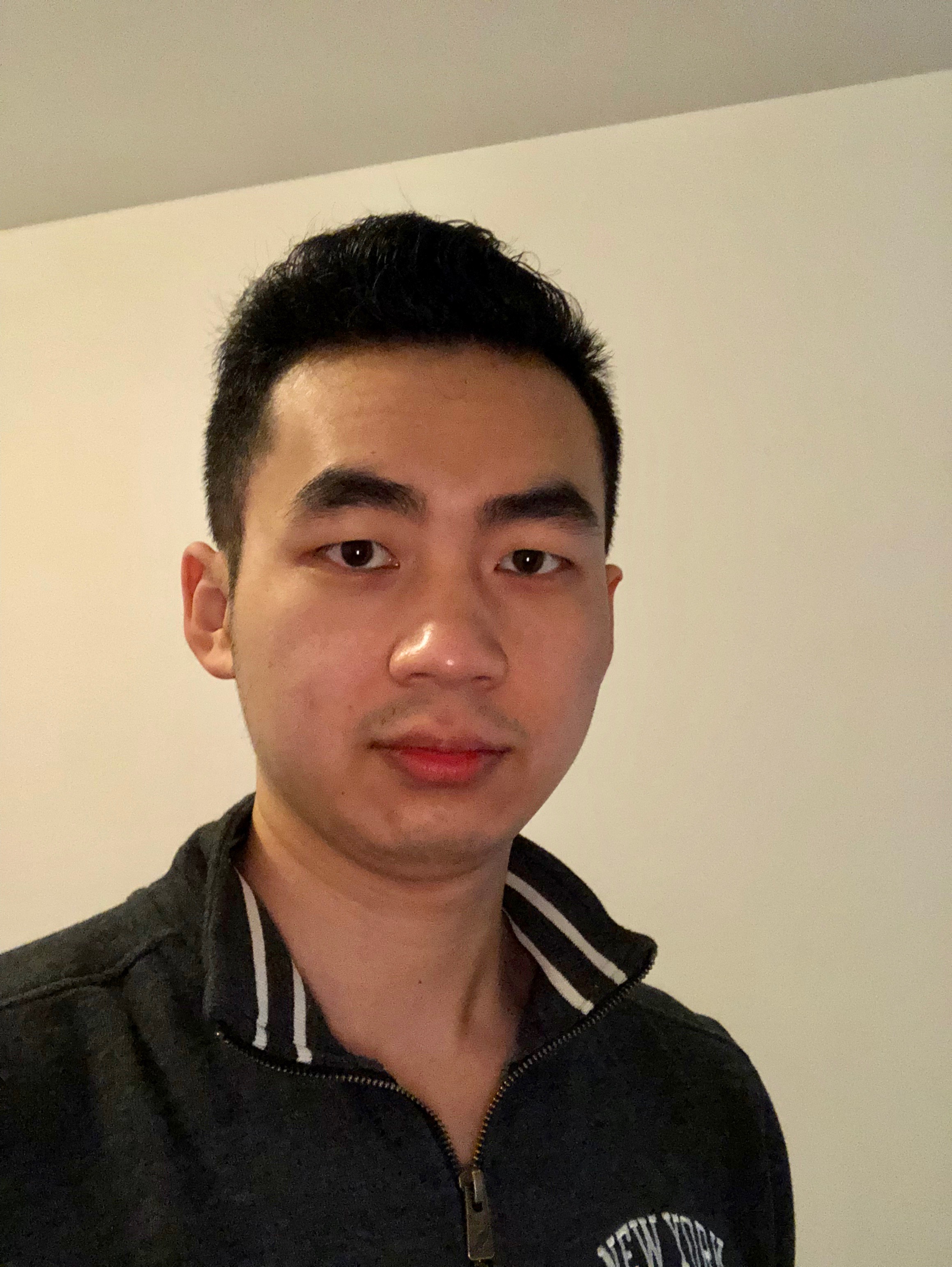}}]{Hao Fu} is a Ph.D. candidate in the Department of Electrical and Computer Engineering at New York University, Tandon School of Engineering, Brooklyn, NY, USA. In 2019, he received his M.S. degree in  the same department. He received his B.S. in Physics from the University of Science and Technology of China, Hefei, China, in 2017.  In 2018 Fall, he joined in Control/Robotics Research Laboratory (CRRL). His research interest is backdooring attacks and security  in cyber-physical systems.

\end{IEEEbiography}

\begin{IEEEbiography}[{\includegraphics[width=1in,height=1.25in,clip,keepaspectratio]{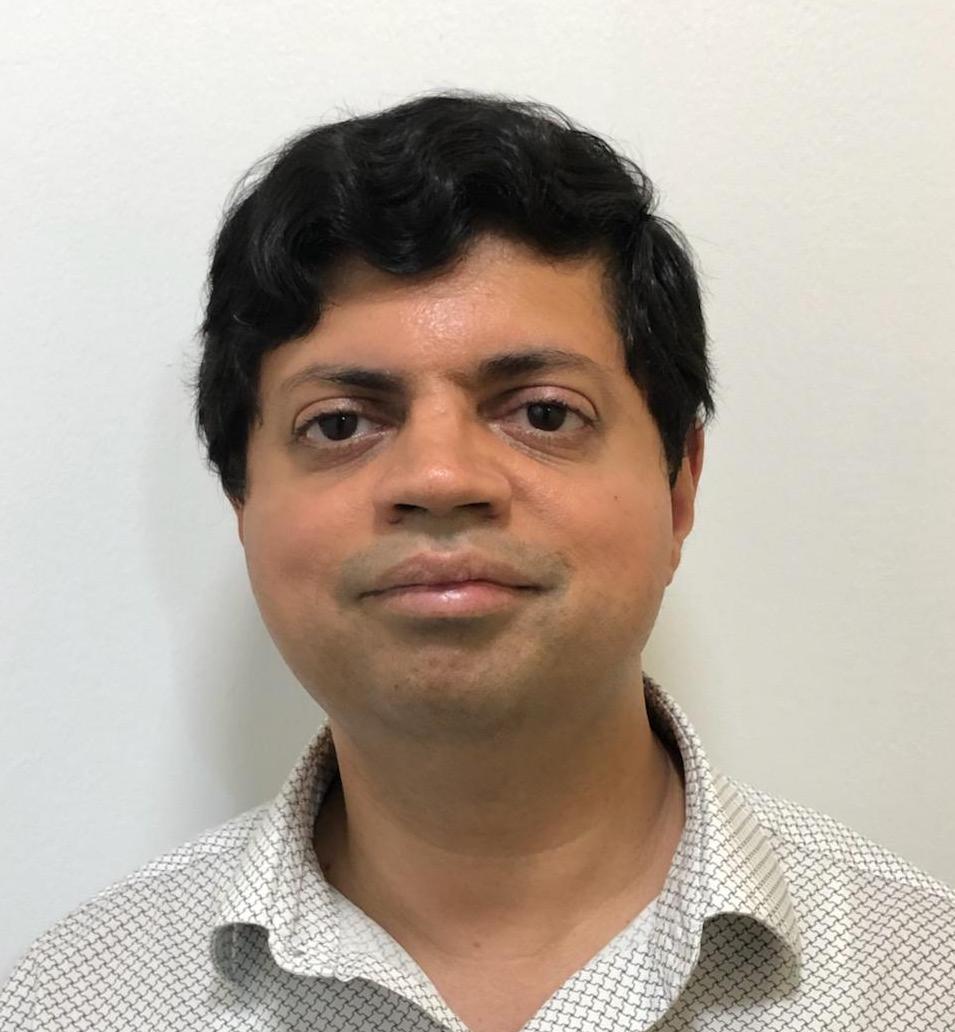}}]{Prashanth Krishnamurthy}
received the B.Tech. degree in electrical engineering from Indian Institute of Technology Madras, Chennai, in 1999, and M.S. and Ph.D. degrees in electrical engineering from Polytechnic University (now NYU) in 2002 and 2006, respectively. He is a Research Scientist and Adjunct Faculty with the Department of Electrical and Computer Engineering at NYU Tandon School of Engineering. He has co-authored over 150 journal and conference papers and a book. His research interests include autonomous vehicles and robotic systems, multi-agent systems, sensor data fusion, robust and adaptive nonlinear control, resilient control, machine learning, real-time embedded systems, cyber-physical systems and cyber-security, decentralized and large-scale systems, and real-time software implementations.
\end{IEEEbiography}

\begin{IEEEbiography}[{\includegraphics[width=1in,height=1.25in,clip,keepaspectratio]{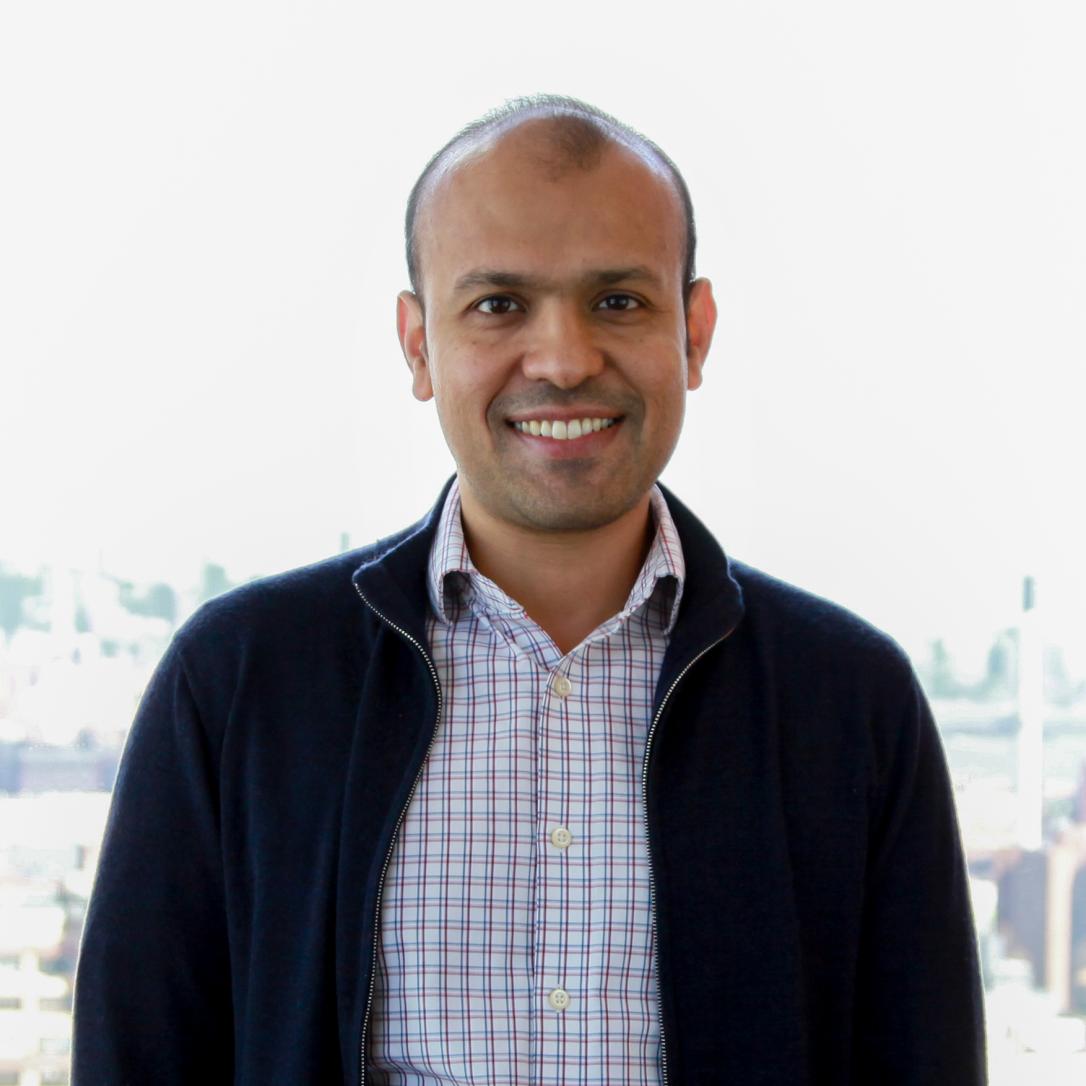}}]{Siddharth Garg}
received the B.Tech. degree in electrical engineering from the Indian Institute of Technology Madras, Chennai, India, and the Ph.D. degree in electrical and computer engineering from Carnegie Mellon University, Pittsburgh, PA, in 2009. He is currently an Associate Professor at New York University, New York, where he joined as an Assistant Professor in 2014. Prior to this, he was an Assistant Professor with the University of Waterloo, Waterloo, ON, Canada, from 2010 to 2014. His current research interests include computer engineering, and more particularly in secure, reliable, and energy efficient computing. He was a recipient of the NSF Career Award in 2015, and the paper awards at the IEEE Symposium on Security and Privacy 2016, the USENIX Security Symposium in 2013, the Semiconductor Research Consortium TECHCON in 2010, and the International Symposium on Quality in Electronic Design in 2009. He was listed in popular science magazine’s annual list of “Brilliant 10” researchers. He serves on the technical program committee of several top conferences in the area of computer engineering and computer hardware and has served as a reviewer for several IEEE and ACM journals conferences.
\end{IEEEbiography}

\begin{IEEEbiography}[{\includegraphics[width=1in,height=1.25in,clip,keepaspectratio]{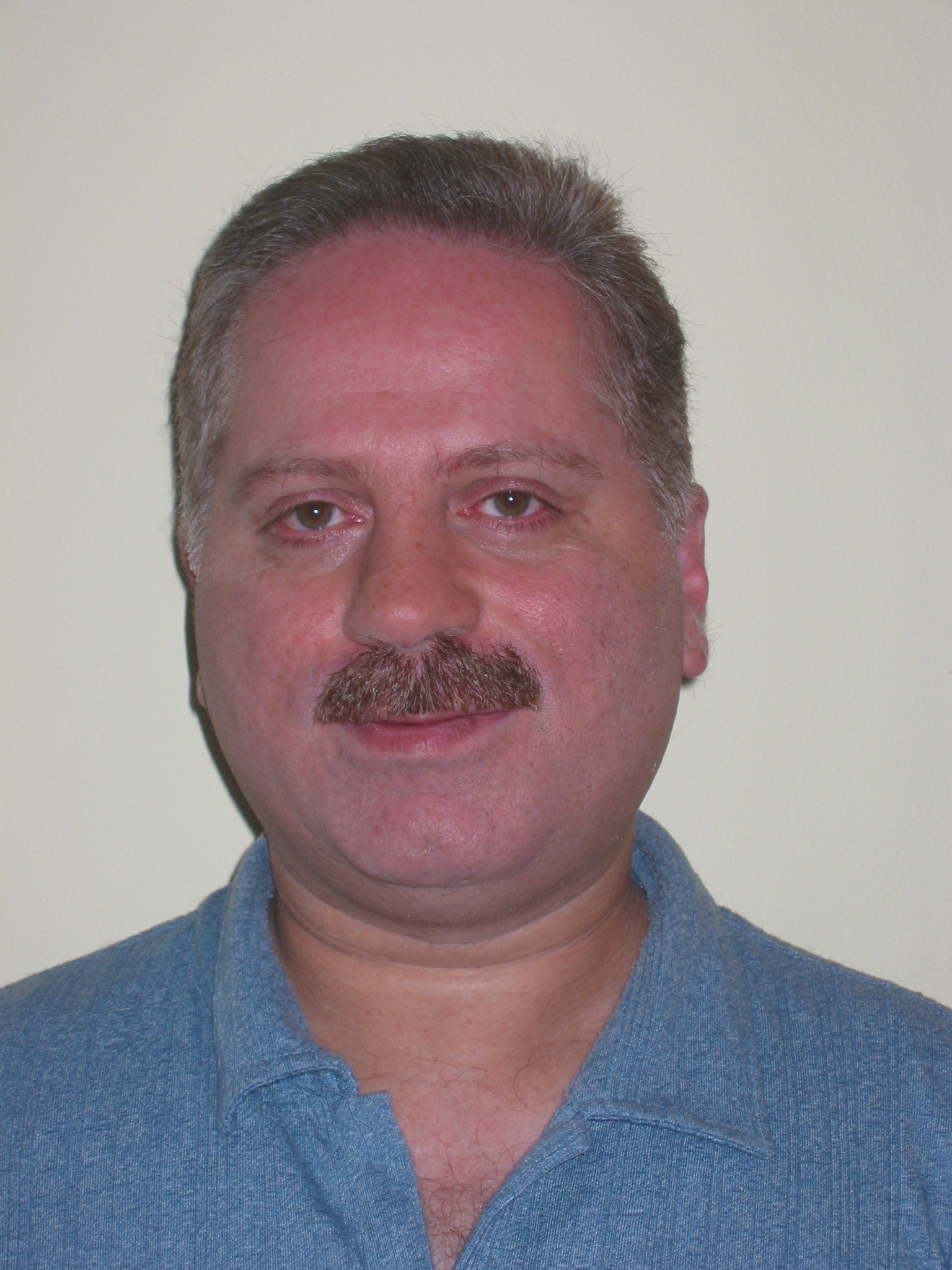}}]{Farshad Khorrami}
received the bachelor’s degrees in mathematics and in electrical engineering from The Ohio State University in 1982 and 1984, respectively, and the master’s degree in mathematics and the Ph.D. degree in electrical engineering from The Ohio State University, in 1984 and 1988, respectively. He is currently a Professor with the Electrical and Computer Engineering Department, NYU, where he joined as an Assistant
Professor in September 1988. He has developed and directed the Control/Robotics Research Laboratory, Polytechnic University (Now NYU) and Co-Director of the Center for AI and Robotics. His research has been supported by the DARPA, ARO, NSF, ONR, DOE, AFRL, NASA, and several corporations. He has published more than 320 refereed journal articles and conference papers in these areas. His book on "modeling and adaptive nonlinear control of electric motors" was published by Springer Verlag in 2003. He also holds 14 U.S. patents on novel smart micropositioners and actuators, control systems, cyber security, and wireless sensors and actuators.  His research interests include adaptive and nonlinear controls, robotics, unmanned vehicles (fixed-wing and rotary wing aircrafts as well as underwater vehicles and surface ships), machine learning, resilient control and cyber security for cyber-physical systems, large-scale systems, decentralized control, and real-time embedded instrumentation and control. He has served as the general chair and a conference organizing committee member for several international
conferences.
\end{IEEEbiography}

\end{document}